\documentclass[11pt]{article}
\pdfoutput=1
\usepackage[utf8]{inputenc}
\usepackage{dsfont}
\usepackage{amsfonts}
\usepackage{amsmath}
\allowdisplaybreaks[4]        
\usepackage{amssymb}
\usepackage{euscript}  
\usepackage{IEEEtrantools}   
\usepackage{braket}
\usepackage{starfont}
\usepackage{color,soul}         
\usepackage{tensor}        
\usepackage{amsthm}
\usepackage{graphicx}
\usepackage{slashed}
\usepackage{leftidx}
\usepackage{subfigure}
\usepackage{bbm}
\definecolor{outerspace}{rgb}{0.25, 0.29, 0.3}
\definecolor{scarlet}{rgb}{1.0, 0.13, 0.0}
\usepackage[header,title,page,titletoc]{appendix}  
\definecolor{princetonorange}{rgb}{1.0, 0.56, 0.0}
\definecolor{WildStrawberry}{rgb}{1.0, 0.26, 0.64}
\definecolor{rossocorsa}{rgb}{0.83, 0.0, 0.0}
\definecolor{navyblue}{rgb}{0.0, 0.0, 0.5}
\usepackage[numbers,sort&compress]{natbib}  
\usepackage{float}
\usepackage[paper=letterpaper,margin=1in]{geometry}
\newcommand{\req}[1]{(\ref{#1})} 

\newcommand{\bea}{\begin{eqnarray}}
\newcommand{\diff}{\mathrm{d}}
\newcommand{\eea}{\end{eqnarray}}
\newcommand{\ba}{\begin{eqnarray}}
\newcommand{\ea}{\end{eqnarray}}

\newcommand{\be}{\begin{equation}}
\newcommand{\ee}{\end{equation} }
\newcommand{\beqa}{\begin{eqnarray}}
\newcommand{\eeqa}{\end{eqnarray}}
\newcommand{\beqar}{\begin{eqnarray*}}
\newcommand{\eeqar}{\end{eqnarray*}}

\renewcommand{\req}[1]{eq.~(\ref{#1})}
\newcommand{\rp}{r_+}

\newcommand{\ssc}{\scriptscriptstyle}
\newcommand{\eg}{{\it e.g.,}\ }
\newcommand{\ie}{{\it i.e.,}\ }

\newcommand{\yp}{y_+}
\newcommand{\xp}{x_+}

\usepackage{hyperref}
\hypersetup{
    colorlinks,
    citecolor=rossocorsa,
    filecolor=navyblue,
    linkcolor=navyblue,
    urlcolor=navyblue
}

\begin{document} 

\begin{titlepage}

\begin{center}

\phantom{ }
\vspace{3cm}

{\bf \Large{Generalized quasi-topological gravities: the whole shebang}}
\vskip 0.5cm
Pablo Bueno,${}^{\text{\Zeus}}$ Pablo A. Cano,${}^{\text{\Kronos}}$ Robie A. Hennigar,${}^{\text{\Apollon}}$ Mengqi Lu,${}^{\text{\Vulkanus}}$ and Javier Moreno${}^{\text{\Hades},\text{\Kronos}}$ 
\vskip 0.05in
\small{${}^{\text{\Zeus}}$ \textit{CERN, Theoretical Physics Department,
}}
\vskip -.4cm
\small{\textit{CH-1211 Geneva 23, Switzerland}}

\small{${}^{\text{\Kronos}}$ \textit{Instituut voor Theoretische Fysica, KU Leuven,}}
\vskip -.4cm
\small{\textit{ Celestijnenlaan 200D, B-3001 Leuven, Belgium}}

\small{${}^{\text{\Apollon}}$  \textit{Departament de F{\'\i}sica Qu\`antica i Astrof\'{\i}sica, Institut de
Ci\`encies del Cosmos,
\vskip -.4cm
 Universitat de
Barcelona, Mart\'{\i} i Franqu\`es 1, E-08028 Barcelona, Spain}}

\small{${}^{\text{\Vulkanus}}$ \textit{Department of Physics and Astronomy, University of Waterloo,}}
\vskip -.4cm
\small{\textit{Waterloo, Ontario, Canada, N2L 3G1}}

\small{${}^{\text{\Hades}}$ \textit{Instituto de F\'isica, Pontificia Universidad Cat\'olica de Valpara\'iso,}}
\vskip -.4cm
\small{\textit{ Casilla 4059, Valpara\'iso, Chile}}

\begin{abstract}




Generalized quasi-topological gravities (GQTGs) are higher-curvature  extensions  of Einstein gravity in $D$-dimensions. Their defining properties include possessing second-order linearized equations of motion around maximally symmetric backgrounds as well as non-hairy generalizations of Schwarzschild's black hole characterized by a single function, $f(r)\equiv -g_{tt}=g_{rr}^{-1}$, which satisfies a second-order differential equation. In \href{https://arxiv.org/abs/1909.07983}{{\tt arXiv:1909.07983}} GQTGs were shown to exist at all orders in curvature and for general $D$. In this paper we prove that, in fact, $n-1$ inequivalent classes of order-$n$ GQTGs exist for $D\geq 5$. Amongst these, we show that one ---and only one--- type of densities is of the Quasi-topological kind, namely, such that the equation for $f(r)$ is algebraic. Our arguments do not work for $D=4$, in which case there seems to be a single unique GQT density at each order which is not of the Quasi-topological kind. We compute the thermodynamic charges of the most general $D$-dimensional order-$n$ GQTG, verify that they satisfy the first law and provide evidence that  they can be entirely written in terms of the embedding function which determines the maximally symmetric vacua of the theory.

\end{abstract}
\end{center}
\end{titlepage}

\setcounter{tocdepth}{2}

{\parskip = .2\baselineskip \tableofcontents}


\section{Introduction}

Higher-curvature theories of gravity play an important role in theoretical physics. On one hand, higher derivatives seem necessary to obtain a consistent quantum description of gravity. For example, string theory and effective field theory approaches predict an infinite tower of higher-curvature corrections to the usual Einstein-Hilbert action~\cite{Callan:1985ia,Zwiebach:1985uq,Bergshoeff:1989de,Gross:1986iv, Gross:1986mw}, while other approaches introduce a finite number of higher-curvature terms to restore certain desirable properties, \eg renormalizability~\cite{Stelle:1976gc, Stelle:1977ry, Starobinsky:1980te}. On the other hand, studying higher-curvature theories can provide insight on the special or universal properties of gravitational theory. This program has been especially fruitful in the holographic context, where deformations of the gravitational theory correspond to deformations of the dual CFT. In this way, it has been possible to provide evidence for universal relationships that hold within holography and beyond~\cite{Myers:2010tj, Myers:2010xs, Perlmutter:2013gua, Mezei:2014zla, Bueno1, Bueno3, Chu:2016tps, Bueno:2018yzo, Li:2018drw, Bueno:2020odt,Bueno:2022jbl}.

In this work we are specifically interested in the structural aspects of a class of theories known as \textit{generalized quasi-topological gravities} (GQTGs). Schematically, we write the action of these theories as
\begin{equation}\label{action}
  S=\frac{1}{16\pi G} \int \diff^D x \sqrt{|g|} \left[\frac{(D-1)(D-2)}{L^2}+R+\sum_{n=2}\sum_{i_n}L^{2(n-1)} \mu_{i_n}^{(n)} \mathcal{R}_{i_n}^{(n)} \right]\, ,
  \end{equation}
where  $\mathcal{R}_{i_n}^{(n)}$ are densities constructed from $n$ Riemann tensors and the metric, the $\mu_{i_n}$ are dimensionless couplings, $L$ is some length scale, and $i_n$ is an index running over all independent GQTG invariants of order $n$. For this action, the field equations can be expressed as
\be\label{EOM} 
\mathcal{E}_{ab} = P_{a}{}^{cde}R_{bcde} - \frac{1}{2} g_{ab} \mathcal{L} - 2 \nabla^c \nabla^d P_{a c d b} = 0\, , \quad \text{with} \quad 
P^{abcd} \equiv \frac{\partial \mathcal{L}}{\partial R_{abcd}}  \, ,
\ee
where $\mathcal{L}$ is the Lagrangian of the theory. For a general theory polynomial in curvature tensors, it is clear that the field equations can contain forth-order derivatives of the metric. The defining property of GQTGs is that they allow for spherically symmetric solutions of the Schwarzschild-like form characterized by a single function  \ie with $g_{tt} g_{rr}=-1$,  where $f(r)$ satisfies at most a second-order equation.. Then, the static spherically symmetric black holes of the theory have  the form
\begin{equation}\label{fEq}
\diff s^2_{f}=-f(r)\diff t^2+\frac{\diff r^2}{f(r)}+r^2\diff \Omega^2_{(D-2)}\, ,
\end{equation}
with $f(r)$ satisfying an equation that contains at most second derivatives.

GQTGs can be further subdivided into different classes depending on the character of the field equations on spherically symmetric and other backgrounds. The most important subclass being Lovelock gravity~\cite{Lovelock1, Lovelock2}, for which the equations on spherically symmetric backgrounds are algebraic in the metric function $f(r)$, and second-order for any metric. Lovelock gravities are also the most constrained. Besides Einstein gravity, there exists no Lovelock theory in $D = 4$, and in general a Lovelock theory of order $n$ in curvature is non-trivial only when $D \ge 2n+1$. A second subclass of the GQTG family are quasi-topological gravities~\cite{Quasi2, Quasi, Dehghani:2011vu,Ahmed:2017jod,Cisterna:2017umf}. For quasi-topological gravities, the field equations for spherically symmetric black holes are algebraic, as for Lovelock theory. However, on general backgrounds the equations of motion will be fourth-order. Quasi-toplogical gravities are less constrained in the sense that they exist in any spacetime dimension $D \ge 5$ for any order in curvature cubic or higher, as explicitly constructed in~\cite{Bueno:2019ycr}. These possibilities do not fully exhaust the space of possible theories, and there exist remaining GQTGs for which the field equations for spherically symmetric black holes is a second-order differential equation for $f(r)$~\cite{PabloPablo, Hennigar:2016gkm, PabloPablo2, Hennigar:2017ego,PabloPablo3} ---these theories can exist even in $D = 4$. 

GQTGs have by now been the subject of quite intensive investigation, \eg~\cite{PabloPablo,Hennigar:2016gkm,PabloPablo2,Hennigar:2017ego,PabloPablo3,Ahmed:2017jod,PabloPablo4,Dey:2016pei,Feng:2017tev,Hennigar:2017umz,Hennigar:2018hza,Bueno:2018xqc,Bueno:2018yzo,Bueno:2018uoy,Poshteh:2018wqy,Mir:2019ecg,Mir:2019rik,Arciniega:2018fxj,Cisterna:2018tgx,Arciniega:2018tnn,Mehdizadeh:2019qvc,Erices:2019mkd,Emond:2019crr,Jiang:2019fpz, Cano:2019ozf, Burger:2019wkq, Bueno:2019ltp,Bueno:2020odt, Frassino:2020zuv, KordZangeneh:2020qeg, Marciu:2020ysf, Quiros:2020uhr, Pookkillath:2020iqq, Marciu:2020ski, Adair:2020vso, Khodabakhshi:2020hny, Edelstein:2020nhg, Konoplya:2020jgt, Khan:2020kwl, Khodabakhshi:2020ddv, Cano:2020qhy, Cano:2020ezi, Quiros:2020eim, Edelstein:2020lgv, Jimenez:2020gbw, Caceres:2020jrf, Mustafa:2020qjo, Cano:2020oaa, Fierro:2020wps, Bhattacharjee:2021nfx, Bhattacharjee:2021jwm, Jawad:2021kkp, Bueno:2021krl, Li:2021jfh, Ghosh:2021zpb, Bakhtiarizadeh:2021vdo, Sardar:2021blt, Bakhtiarizadeh:2021hjr, Gray:2021roq, Jaime:2022cho, Bueno:2022lhf, Edelstein:2022xlb, Cano:2022ord}, and many of the interesting properties of these theories are now well-understood. Here we summarize some particularly relevant ones:
\begin{enumerate}
\item When linearized around any maximally symmetric background, their equations are identical to the Einstein gravity ones, up to a redefinition of the Newton constant ---in other words, they only propagate the usual transverse and traceless graviton in the vacuum \cite{PabloPablo,Hennigar:2016gkm,PabloPablo2,Hennigar:2017ego,PabloPablo3,Ahmed:2017jod,PabloPablo4}.
\item They possess non-hairy black hole solutions fully characterized by their ADM mass/energy and whose thermodynamic properties can be obtained from an algebraic system of equations.
\item Although the defining property pertains to static spherically symmetric black holes, certain subsets of GQTGs allow for reduction of order in the field equations for other metrics, such as Taub-NUT/Bolt~ \cite{Bueno:2018uoy}, slowly-rotating black holes~\cite{Adair:2020vso, Gray:2021roq}, near extremal black holes~\cite{Cano:2019ozf}, and cosmological solutions~\cite{Arciniega:2018fxj, Arciniega:2018tnn,Cisterna:2018tgx, Cano:2020oaa}.
\item In the context of gravitational effective field theory, any higher-curvature theory can be mapped, via field redefinition, into some GQTG~\cite{Bueno:2019ltp, Bueno:2019ycr}.
\item We can consider arbitrary linear combinations of GQTG densities and the corresponding properties hold, which means, in particular, that GQTG theories have a well-defined and continuous Einstein gravity limit, corresponding to setting all higher-curvature couplings to zero.
\item Extensions away from pure metric theories, including scalars or vector fields, while preserving the main properties are possible~\cite{Cano:2020qhy,Bueno:2021krl,Cano:2022ord}.
\end{enumerate}

Our purpose here is to complete the study of structural aspects of GQTGs. In~\cite{Bueno:2019ycr} we proved existence of GQTGs at all orders of curvature and in all dimensions $D \ge 4$. In this manuscript, we will address how many \textit{distinct/inequivalent} GQTGs exist at each order in curvature and in each dimension. The organization of the manuscript is as follows. We begin in Section~\ref{GQTGss} by reviewing in more detail the defining properties of GQTGs and introducing notation that will be used throughout. Then, in Section~\ref{GQTGssss}, we provide a simple argument that gives an upper bound on the number of possible distinct GQTGs. We then refine this upper bound into an exact result, showing that at order $n$ in curvature there are $n-1$ distinct GQTGs, provided $D > 4$, while there is a single unique family provided $D = 4$. In Section~\ref{GQTGthermo} we compute the thermodynamic charges for any possible GQTG, and verify that they satisfy the first law. Intriguingly, we find good evidence that the thermodynamics for GQTGs can be written entirely in terms of the embedding function for the given family of theories, which determines the maximally symmetric vacua of the theory. We collect a number of useful results and expressions in the appendix.

\section{Generalized quasi-topological gravities}\label{GQTGss}
We start in this section with a quick review of the defining properties of GQTGs and some notation. Our discussion here closely follows that of~\cite{Bueno:2019ycr}. We also introduce the notion of {\it inequivalent} GQTG densities which will be important for the rest of the paper. Roughly speaking, we will say that two GQTG densities of a fixed curvature order $n$ are inequivalent if they give rise to different equations for the metric function $f(r)$.

\subsection{Definitions}
A general static and spherically symmetric (SSS) metric can be written in terms of two undetermined functions $N(r)$ and $f(r)$ as
\begin{equation}
\label{SSS}
\diff s^2_{N, f}=-N(r)^2 f(r) \diff t^2+\frac{\diff  r^2}{f(r)}+ r^2\diff \Omega^2_{(D-2)}\, ,
\end{equation}
where $\diff \Omega^2_{(D-2)}$ is the metric of the $(D-2)$-dimensional round sphere. Essentially all our discussion extends straightforwardly to the cases in which the horizon is planar or hyperbolic instead. The formulas below will include those as well, the different cases being parametrized by a constant $k$ taking values $k=1,0,-1$ for spherical, planar and hyperbolic horizons respectively.  

For a given curvature invariant of order $n$, $\mathcal{R}_{(n)}$, we define $L_{N,f}$ and $S_{N,f} $  as the effective Lagrangian and on-shell action which result from evaluating $\sqrt{|g|}\mathcal{R}_{(n)}$ in the ansatz (\ref{SSS}) 
\begin{equation}\label{ansS}
L_{N,f}\equiv \left. N(r) r^{D-2} \mathcal{R}_{(n)}\right|_{N,f}\, , \quad S_{N,f}\equiv \Omega_{(D-2)}\int \diff t  \int  \diff r L_{N,f} \, ,
\end{equation}
where we integrated over the angular directions, $\Omega_{(D-2)}\equiv 2\pi^{\frac{D-1}{2}}/\Gamma[\frac{D-1}{2}]$. We will define $L_f\equiv L_{1,f}$ and $S_f\equiv S_{1,f}$, namely, the expressions obtained from setting $N=1$  in $L_{N,f}$. 
Now, solving the full nonlinear equations of motion for a metric of the form (\ref{SSS}) can be shown to be equivalent to solving  the Euler-Lagrange equations of $S_{N,f}$ associated to $N(r)$ and $f(r)$ \cite{Palais:1979rca,Deser:2003up,PabloPablo4}, namely,
\be
\left.\mathcal{E}^{ab}\right|_{N,f}\equiv \left. \frac{1}{\sqrt{|g|}}  \frac{\delta S}{\delta g^{ab}}  \right|_{N,f}=0 \quad   \Leftrightarrow   \quad \frac{\delta S_{N,f}}{\delta N}= \frac{\delta S_{N,f}}{\delta f}=0\, .
\ee

We say that $\mathcal{R}_{(n)}$ is a GQTG density if the Euler-Lagrange equation of.  $f(r)$ associated to  $L_f$  vanishes identically, \ie if
\begin{equation} \label{GQTGcond}
\frac{\delta S_{f}}{\delta f}=0\, ,   \quad \forall \, \, f(r)\, .
\end{equation}
This is the same as asking $L_f$ to be a total derivative, 
\begin{equation}\label{condd2}
L_f =T_0'\, ,
\end{equation}
for some function  $T_0(r,f(r),f'(r))$.

The equation satisfied by $f(r)$ for a given GQTG density can be obtained from the variation of  $L_{N,f}$ with  respect to $N(r)$ as 
\begin{equation}\label{eqf}
\left.\frac{\delta S_{N,f}}{\delta N}\right|_{N=1}=0\, \quad \Leftrightarrow\quad \text{equation of}\quad f(r)\, .
\end{equation}
As explained in \cite{PabloPablo3}, whenever \req{condd2} holds, the effective Lagrangian  $L_{N,f}$ takes the form
\begin{equation}\label{fofwo}
L_{N,f}=N T_0' +  N' T_1 +  N'' T_2 +\mathcal{O}(N'^2/N)\, ,
\end{equation}
where $T_{1}$, $T_2$  are functions of $f(r)$ and its derivatives, and $\mathcal{O}(N'^2/N)$ is a sum of terms all of which are at least quadratic in derivatives of $N(r)$. Integrating by parts it follows  that 
\begin{equation}
S_{N,f} = \Omega_{(D-2)}  \int \diff t \int \diff r \left[N\left(T_0-T_1+T_2' \right)' +\mathcal{O}(N'^2/N) \right]\, .
\end{equation}
So it is possible to write all terms  involving one power of $N(r)$ or its derivatives as a product of $N(r)$ and a total derivative which depends on $f(r)$ alone. Now, it follows straightforwardly that condition (\ref{eqf}) equates that total derivative to zero. Integrating it once one we are left with \cite{PabloPablo3}
\begin{equation} \label{eqqqf}
\mathcal{F}_{\mathcal{R}_{(n)}}  \equiv T_0-T_1+T_2'=C\, ,
\end{equation}
where $C$ is an integration constant related to the ADM mass of the solution \cite{Arnowitt:1960es,Arnowitt:1960zzc,Arnowitt:1961zz,Deser:2002jk}. In particular, for spherical horizons, the precise relation reads 
\begin{equation}
C= \frac{M}{\Omega_{(D-2)}}\, .
\end{equation}
Hence, given some linear combination of GQTG densities, obtaining the equation satisfied by the metric function $f(r)$ amounts to evaluating $L_{N,f}$ as defined in \req{ansS} and then identifying the functions $T_{i=0,1,2}$ from \req{fofwo}. The equation is then given by (\ref{eqqqf}).\footnote{Sometimes we will refer to this equation as the ``integrated equation'' of $f(r)$ to emphasize the fact that it follows from integrating once (on $r$) the only non-vanishing component of the actual equations of motion of the theory evaluated on the single-function SSS ansatz. }

As argued in \cite{PabloPablo3}, the integrated equation is at most second-order in derivatives of  $f(r)$. In fact, there are two possibilities as far as the number of derivatives of $f(r)$ are involved: i) theories whose integrated equation involves  $f'(r)$ and $f''(r)$; ii) theories whose integrated equation exclusively involves $f(r)$, so the equation is algebraic instead of differential. We shall call theories of the former class ``genuine'' GQTG densities.
 Theories of the latter class are called Quasi-topological gravities, and they include Einstein and Lovelock theories as subcases.

Now, a natural question is: given a fixed spacetime dimension $D$ and a curvature order $n$, are the integrated equations corresponding to different genuine GQTG densities $\{ \mathcal{R}^{I}_{(n)}$,   $\mathcal{R}^{II}_{(n)}$, $\dots$  $\mathcal{R}^{i_n}_{(n)}\}$ proportional to each other ---\ie are the functional dependences on $r$, $f(r)$, $f'(r)$ and $f''(r)$ of the equations  identical--- for the various densities? If not, how many inequivalent contributions to the equation of $f(r)$ are there at a given order in curvature?  Analogous questions can be asked fixing $D$ and $n$ for theories belonging to the Quasi-topological class. Given two genuine GQTG densities of order $n$, we will say they are ``inequivalent'' (as far as SSS solutions are concerned) if the quotient of their integrated equations is not a constant,
\begin{equation}
\mathcal{R}^{I}_{(n)} \quad \text{inequivalent from} \quad \mathcal{R}^{II}_{(n)} \quad \Leftrightarrow \quad  \frac{\mathcal{F}_{\mathcal{R}^I_{(n)}}(r,f(r),f'(r),f''(r)) }{\mathcal{F}_{\mathcal{R}^{II}_{(n)}} \left(r,f(r),f'(r),f''(r)\right)} \neq \text{constant} \, .
\end{equation}
Otherwise we will say they are ``equivalent''.
Given two Quasi-topological gravities of order $n$, we would perform an analogous definition,
 \begin{equation}
\mathcal{Z}^{I}_{(n)} \quad \text{inequivalent from} \quad \mathcal{Z}^{II}_{(n)} \quad \Leftrightarrow \quad  \frac{\mathcal{F}_{\mathcal{Z}^I_{(n)}}(r,f(r)) }{\mathcal{F}_{\mathcal{Z}^{II}_{(n)}} \left(r,f(r)\right)} \neq \text{constant} \, .
\end{equation}
but we will show later that, in fact, all Quasi-topological gravities of a given order are equivalent. That will not be the case for genuine GQTGs, in whose case we will prove that there exist $(n-2)$ inequivalent densities for $D\geq 5$.\footnote{The existence of multiple types of GQTG densities was first pointed out in \cite{Bueno:2020odt}, where two inequivalent quintic densities were explicitly constructed in $D=6$.}



\section{How many types of GQTGs are there?}\label{GQTGssss}
In this section we prove that there exist exactly $(n-2)$ inequivalent genuine GQTG densities and a single inequivalent Quasi-topological one at a given curvature order $n$ in $D\geq 5$. In $D=4$ there are no Quasi-topological theories and we argue that our proof for the existence of  $(n-2)$ genuine GQTG densities fails in that case, illustrating the fact that a single genuine GQTG density exists in $D=4$ for $n\geq 3$. 

\subsection{At most $(n+1)$ order-$n$ densities}
Let us start our study by putting an upper bound on the possible number of inequivalent GQTG densities existing at a given curvature order $n$. As argued in  \cite{Deser:2005pc}, evaluated on a metric of the form (\ref{fEq}), the Riemann tensor can be written as 
\begin{equation}\label{rieef}
\left.\tensor{R}{^{ab}_{cd}}\right|_f=2\left[-A T^{[a}_{[c}T^{b]}_{d]}+2B T^{[a}_{[c}\sigma^{b]}_{d]}+\psi \sigma^{[a}_{[c}\sigma^{b]}_{d]}\right]\, ,
\end{equation}
where $\sigma_a^b$ and $T_a^b$ are projectors on the angular and ($t$,$r$) directions, respectively.\footnote{These satisfy  $T_a^b T_b^c=T_a^c$,  $\sigma_a^b \sigma_b^c=\sigma_a^c$, $\sigma_a^b T_b^c=0$,  $\delta^a_b T_{a}^b=2$, $\delta^a_b \sigma_a^b=(D-2)$, $\delta^{a}_{b}=T^{a}_{b}+\sigma^{a}_{b}$. } On the other hand,  the dependence on the radial   coordinate appears exclusively through the three functions  $A$, $B$ and $\psi$, which read
\begin{equation}
A\equiv \frac{f''(r)}{2}\, ,   \quad   B\equiv -\frac{f'(r)}{2r}\, , \quad \psi\equiv \frac{k-f(r)}{r^2}\, ,
\end{equation}
where  $k={1,0, -1}$ for spherical, planar and  hyperbolic horizons respectively.

Now, GQTG densities are built from contractions of the metric and the Riemann tensor, so any order-$n$ density of that type will become some polynomial of these objects when evaluated on (\ref{fEq}), namely,
\begin{equation}\label{rfff}
\mathcal{S}\big|_f=\sum_{l=0}^n\sum_{k=0}^l c_{k,l}B^l\psi^{l-k}A^{n-l}\, ,
\end{equation}
for some constants $c_{k,l}$. The idea is now to determine the most general constants  $c_{k,l}$ consistent with the GQTG requirement, which asks $r^{D-2} \mathcal{S}|_f $ to be a total derivative, \ie
\begin{equation}\label{pazo}
r^{D-2} \mathcal{S}|_f = T_0'(r)\, .
\end{equation}
Note that imposing this condition on \req{rfff} and finding the compatible values of $c_{k,l}$ does not guarantee that the corresponding GQTG densities actually exist, as this does not provide an explicit construction of covariant curvature densities. Doing this does impose, nonetheless, a necessary condition which all actual densities must satisfy.  
Given a GQTG density, $\mathcal{S}$, it is useful to define the object $\tau(r)$ through the relation
\begin{equation}\label{ttau}
T_0 \equiv r^{D-1} \tau\, , \quad \text{so that} \quad  \mathcal{S}\big|_f= \frac{1}{r^{D-2}} \frac{\diff }{\diff r} \left[ r^{D-1} \tau(r)\right] \, .
\end{equation}
In a sense, $\tau(r)$ is the fundamental building block as long as on-shell GQTG densities are concerned. Observe that since
\begin{equation}\label{}
 \sum_i \alpha_i  \mathcal{S}_i\big|_f = \frac{1}{r^{D-2}} \frac{\diff }{\diff r} \left[ r^{D-1} \sum_i \alpha_i \tau_{(i)}(r)\right] \, ,
\end{equation}
linear combinations of the $\tau_{(i)}$ give rise to linear combinations of GQTG densities in an obvious way.

Now, imposing (\ref{pazo}) on densities of the form \req{rfff}, we find that there are $(n+1)$ independent possible densities at a given order $n$. In terms of the $\tau(r)$, the possibilities turn out to be simply given by $\tau=\tau_{(n,j)}$, where we defined
\begin{equation}
\tau_{(n,j)}\equiv \psi^{n-j} B^j \, , \quad \text{where} \quad j=0,1,\dots,n\, .
\end{equation}
The corresponding putative on-shell densities read\footnote{Note that for the objects $\mathcal{S}_{(n,j)}$ we omit the $|_f$. By this we mean that we literally define $\mathcal{S}_{(n,j)}$ to be the expression that appears in the right-hand side. Actual densities evaluated on the single-function SSS ansatz will reduce to linear combinations of the $\mathcal{S}_{(n,j)}$.  }
\begin{equation}\label{rj}
\mathcal{S}_{(n,j)} \equiv \frac{1}{r^{D-2}} \frac{\diff}{\diff r} \left[ r^{D-1} \tau_{(n,j)}\right] \, , \quad j=0,1,\dots,n\, .
\end{equation}
Observe that the resulting possibilities are such that $A$ only appears either to the power $1$ or to the power $0$ when expanding $\mathcal{S}_{(n,j)}$, which is like restricting the sum in $l$ appearing in (\ref{rfff}) to $l=\{n-1,n\}$.
It follows that any GQTG density in any number of dimensions and at any order in curvature must necessarily be expressible as a linear combination of the above densities when evaluated on the single-function SSS ansatz, namely
\begin{equation}\label{genS}
\mathcal{S}|_f=  \frac{1}{r^{D-2}} \frac{\diff}{\diff r} \left[ r^{D-1} \sum_{j=0}^n \alpha_{(n,j)} \tau_{(n,j)}(r)\right] \, ,
\end{equation}
for certain constants $\alpha_{(n,j)}$. 

Using the methods developed in~\cite{Bueno:2019ycr} ---cf. section 5 of that work--- it is possible to compute the field equations for the putative theory~\eqref{genS} despite the fact that a covariant form of the action is not known. The integrated equation for the metric function $f(r)$ corresponding to a putative density $\mathcal{S}_{(n,j)}$ is given, in the notation of \req{eqqqf}, by\footnote{So, for a linear combination of densities, the equation would read $\sum_j \alpha_{(n,j)} \mathcal{F}_{(j)}=C $ where $C$ is an integration constant related to the mass of the solution.}
\begin{align} \label{fnj}
&\mathcal{F}_{(n,j)}=  \frac{(-1)^{j+1}}{2^{j+1}}r^{D-2+j-2n} (k-f)^{n-j-1}(f')^{j-2} \times \\ &\left[ f' \Big[j(D-1+j-2n)(k-f)f-(j-1)r (k+(n-j-1)f) f' \right] +j(j-1)  r (k-f) f f''\Big] \, . \notag
\end{align}
Observe that this simplifies considerably both for $j=0$ and $j=1$. In those cases the dependence on $f'$ and $f''$ disappears and one finds algebraic equations for $f(r)$,
\begin{equation}
\mathcal{F}_{(n,0)}=  -\frac{r^{D-1-2n}}{2} (k-f)^{n-1} [k+(n-1)f]\, , \quad \mathcal{F}_{(n,1)}=\frac{(D-2n)r^{D-1-2n}}{4}(k-f)^{n-1}f \, .
\end{equation}
An obvious question at this point is: which of these possible densities actually corresponds to the Einstein-Hilbert one, if any. In that case we have $n=1$, and the two possible densities and their integrated equations of motion read, respectively,
\begin{align}
\mathcal{S}_{(1,0)}&=-\frac{1}{r^2} \left[(D-3) (f-k)+r f' \right]\, , \quad \mathcal{F}_{(1,0)}=-\frac{r^{D-3} k }{2} \, , \\
\mathcal{S}_{(1,1)}&=-\frac{1}{2r^2} \left[(D-2)r f'+r^2 f'' \right]\, , \quad \mathcal{F}_{(1,1)}=\frac{(D-2)r^{D-3}f }{4} \, .
\end{align}
Now, the corresponding expressions for the Einstein-Hilbert action (\ie for a density given by the Ricci scalar $\mathcal{S}_{\rm \ssc EH}\equiv R$) read
\begin{equation}
\mathcal{S}_{\rm \ssc EH}|_f=-\frac{1}{r^2} \left[(D-2)(D-3)(f-k)+ 2(D-2)r f'+r^2 f''\right]\, , \quad \mathcal{F}_{\rm \ssc EH}=-(D-2)(f-k) r^{D-3} \, .
\end{equation}
Hence, none of the putative densities coincides with the Einstein-Hilbert one. Rather, it is a linear combination of the two which does, namely,
\begin{equation}
\mathcal{S}_{\rm \ssc EH}|_f= (D-2) \mathcal{S}_{(1,0)} + 2 \mathcal{S}_{(1,1)}\, .
\end{equation}
Even though our approach has selected two possible independent densities susceptible of giving rise to GQTG densities at linear order in curvature, there (obviously) exists a unique possibility corresponding to an actual density, given by the Ricci scalar, which therefore is given by a linear combination of the two. While the $n=1$ case is somewhat special, this already illustrates the fact that our upper bound of $(n+1)$ densities at order $n$ is not tight and can be improved. For higher $n$, the only known examples of densities which give rise to algebraic integrated equations for $f(r)$ are Lovelock and Quasi-topological gravities. From our perspective, at a given order $n$ in $D$ dimensions, all available Lovelock and Quasi-topological gravities for such $n$ and $D$ are ``equivalent'' as far as the equation of $f(r)$ is concerned, which means that they should correspond to a fixed linear combination of $\mathcal{S}_{(n,0)}$ and $\mathcal{S}_{(n,1)}$. In the next subsections we argue that, indeed, the bound of $(n+1)$ densities can be lowered to at most $(n-1)$  GQTG densities of order $n\geq 2$. While amongst the $(n+1)$ candidates identified here there are two which produce algebraic equations, we will see that only a linear combination of the two survives, precisely corresponding to the known Lovelock and Quasi-topological case. The additional putative $(n-2)$ densities would give rise to distinct second-order differential equations for $f(r)$.



\subsection{At most $(n-1)$ order-$n$ densities}\label{atmostnm1}
In order to lower our upper bound on the number of available GQTG densities existing at a given order, we can impose some further conditions on our candidate on-shell densities $\mathcal{S}_{(n,j)}$. The first condition comes from imposing that the equations of motion associated to them admit maximally symmetric solutions. When evaluated on such backgrounds, the equations of motion of actual higher-curvature densities reduce to an algebraic equation which involves the cosmological constant, the curvature scale of the background (\eg the AdS radius) as well as the higher-curvature couplings. More precisely, consider a gravitational Lagrangian consisting of a linear combination of generic higher-curvature densities of the form given in \req{action}.
The result for the equations of motion when evaluated for
\begin{equation}
f(r)=\frac{r^2}{L_{\star}^2}+k\, ,
\end{equation}
which corresponds to pure AdS$_{D}$ with radius $L_{\star}$, is given by
\begin{equation}
\frac{r^{D-1}}{16\pi G} \left[\frac{(D-2)}{L^2}-\frac{(D-2)}{L_{\star}^2}+ \sum_{n=2} \sum_{i_n} \frac{L^{2(n-1)}}{L_{\star}^{2n}}  \mu_{i_n}^{(n)} a_{i_n}^{(n)}\right]=0\, , 
\end{equation}
for certain constants $a_{i_n}^{(n)}$. Interestingly, as we will see below, this same equation which determines the vacua, also appears to play a key role in the thermodynamics of black holes in the theory. Naturally, the solution for Einstein gravity is simply $L^2=L_{\star}^2$, which relates the action scale to the AdS radius in the usual way.

Now, what happens when we consider the integrated equations of a linear combination of candidate on-shell GQTG densities, each contributing as in \req{fnj}, on such a background? It turns out that the result $\sum_j \alpha_{(n,j)} \mathcal{F}_{(n,j)}$ contains two different kinds of terms, one which goes with a power of $r^{D-1}$, and one which goes with a power of $r^{D-3}$. As we have seen, actual densities contribute with a single power of the type  $r^{D-1}$, so we must impose that the second kind of term is absent for our putative densities. Removing such a piece amounts to imposing the condition
\begin{equation}\label{cond1}
\sum_{j=0}^{n} \alpha_{(n,j)} (2n-Dj)=0\, .
\end{equation}
Hence, we learn that not all the candidate densities can be independent and we reduce the number from $(n+1)$ to $n$. 

There is another condition we can easily impose on our candidate densities. As explained in the first section, GQTG densities have second-order linearized equations around general maximally symmetric backgrounds. This is in contradistinction to most higher-curvature gravities, whose linearized equations involve up to four derivatives of the metric ---see \eg \cite{Aspects} for general formulas. Suppose then that we consider a small radial perturbation on AdS space such that the metric function becomes
\begin{equation}\label{perv}
f(r)=\frac{r^2}{L_{\star}^2}+k+\varepsilon h(r)\, ,
\end{equation}
where $\varepsilon \ll 1$. Now, observe that in our general discussion, the integrated equation of motion for a GQTG density, $\mathcal{F}_{\mathcal{S}_n}$, has been integrated once (on $r$) with respect to the actual equations of motion of the corresponding density. Hence, the fact that the actual (linearized) equations of motion for GQTG densities are second order for any perturbation on a maximally symmetric background implies that the integrated equations cannot contain terms involving $h''(r)$ (or more derivatives) at leading order in $\varepsilon$. If they did, the actual linearized equations would involve terms of the form $\sim \varepsilon h'''(r)$, in contradiction with the linearized second-order behavior. With this in mind, our strategy now is to insert \req{perv} in a linear combination of integrated equations for our candidate on-shell densities (\ref{fnj}) and impose that no terms involving $h''(r)$ appear at leading order in $\varepsilon$. By doing so, we find an additional (remarkably simple) condition, which reads
\begin{equation}\label{cond2}
\sum_{j=0}^{n} \alpha_{(n,j)}j (j-1)=0\, .
\end{equation}
Imposing it further reduces the number of independent densities from $n$ to $(n-1)$. Hence, we conclude that in $D$ dimensions there exist at most $(n-1)$ inequivalent GQTG theories of order $n$. Later in subsection \ref{proofexa} we will prove that in fact there exist exactly $(n-1)$ inequivalent densities for $D\geq 5$. There are many possible ways to choose a basis of on-shell densities so that \req{cond1} and \req{cond2} are implemented. For instance, we may choose for the $\tau(r)$ functions defined in \req{ttau}
\begin{align}\label{qt}
\tau^{\rm QT}_{\{n\}}\equiv & +(2n-D) \tau_{(n,0)}-2n \tau_{(n,1)}\, ,\\ 
\tau^{\rm GQT}_{\{n,j\}}\equiv& +(j+1)(D j -4 n) \tau_{(n,j+1)} \\ \notag &+ \left[2D(1-j^2)-4n(1-2j) \right]\tau_{(n,j)} + j [D(j+1)-4n]\tau_{(n,j-1)}\, ,
\end{align}
with $j=2,\dots,n-1$, where we isolated the QT class combination in the first line ---see next subsection.

Naturally, constructing actual covariant densities of each of the classes is a non-trivial problem on its own. Explicit formulas for order-$n$ GQTG densities in arbitrary dimensions $D\geq 4$ as well as for order-$n$ QT densities in $D\geq 5$ were presented in \cite{Bueno:2019ycr}. However, these cases only exhausted $2$ of the $(n-1)$ classes which we show to exist for $D\geq 5$ in the present paper (one of the genuine GQTG types and the Quasi-topological one). In Appendix \ref{densiii} we present explicit formulas for the $(n-2)$ different types of GQTG densities for $n=4,5,6$ in $D=5$ and $D=6$.

\subsection{Uniqueness of Quasi-topological densities}
As mentioned above, Quasi-topological densities are a subclass of GQTGs characterized by having an algebraic (as opposed to second-order differential) integrated equation of motion for the metric function $f(r)$ \cite{Quasi2, Quasi, Dehghani:2011vu,Ahmed:2017jod,Cisterna:2017umf}. Theories of that kind are required to satisfy an additional condition besides  
(\ref{GQTGcond}), namely \cite{Bueno:2019ycr} 
\begin{equation}
\left[ \frac{D-2}{r} \frac{\partial}{\partial f''}+\frac{\diff }{\diff  r} \frac{\partial}{\partial f''}+\frac{(D-3)}{2} \frac{\partial}{\partial f'}+\frac{r}{2} \frac{\diff }{\diff  r} \frac{\partial}{\partial f'}-r \frac{\partial}{\partial f}\right] \mathcal{Z}|_f=0\, ,
\end{equation}
which is equivalent to enforcing that the term $\nabla^d P_{acdb}$ from the field equations vanishes on a static spherically symmetric metric ansatz.
Imposing this condition on a general linear combination of our canditate densities  (\ref{genS}) severely constrains the values of the $\alpha_j$, and we find that $\tau^{\rm QT}_{\{n\}}$ as defined in \req{qt}
is in fact the only possibility. Hence, we learn that the only combination of putative densities compatible with the Quati-topological condition is given by
\begin{equation}\label{zf}
\mathcal{Z}_{(n)}|_f= \frac{1}{r^{D-2}} \frac{\diff}{\diff r} \left[ r^{D-1} \left( (2n-D) \tau_{(n,0)}-2n \tau_{(n,1)}\right) \right]\, .
\end{equation}
Now, Quasi-topological gravities with precisely this structure were shown to exist in \cite{Bueno:2019ycr} at all orders in $n$ and for all $D\geq 5$. Therefore, we conclude that the only possible on-shell structure of a Quasi-topological density is given by (\ref{zf}). There are no additional inequivalent Quasi-topological densities besides the known ones: if a given higher-curvature density possesses second-order linearized equations around maximally symmetric backgrounds and admits black hole solutions satisfying $g_{tt}g_{rr}=-1$ and such that the equation for $f(r)$ is algebraic, then, the equation which determines such a function is uniquely determined to be
\begin{equation}
\mathcal{F}_{\mathcal{Z}_n}=\frac{(D-2n)}{2} r^{D-2n-1} (k-f)^n\, .
\end{equation}
This naturally includes the subcases of Einstein and Lovelock gravities. 

\subsection{Exactly $(n-1)$ order-$n$ densities}\label{proofexa}
Let us finally proceed to prove that there exist exactly $(n-1)$ inequivalent GQTG densities of order $n$ in dimensions higher than four.

Consider the following combination of ``on-shell densities''
\begin{equation}
\mathcal{S}^{(k)}_{p}=\sum_{i=0}^{p}\alpha^{(k)}_{p,i}  \mathcal{S}_{(p,i)}\, ,\quad k=1,\ldots,k _p\equiv \max\,(1,p-1)\, ,
\end{equation}
where the $\mathcal{S}_{(p,i)}$ are defined in \req{rj} and where we assume the constants $\alpha^{(k)}_{p,i}$ to satisfy the constraints found in subsection \ref{atmostnm1}, namely,
\begin{equation}\label{alphaconstrv2}
\sum_{j=0}^{p} \alpha^{(k)}_{p,j} (2p - D j )=0\, ,\quad \sum_{j=0}^{p} \alpha^{(k)}_{p,j} j(j-1)=0\, .
\end{equation}
At each curvature order $p$, there are $k_{p}$ linearly independent solutions and the index $k$ labels each of them.  

Now, let us assume that for $p=1,2,\ldots ,n$ we have proven that all of these on-shell densities correspond to the evaluation of actual higher-curvature densities on the single-function SSS ansatz.  Namely, there exists a set of Lagrangians $\mathcal{R}_{p}^{(k)}$ such that 
\begin{equation}
\mathcal{R}_{p}^{(k)}\Big|_{f}=\mathcal{S}^{(k)}_{p}\, ,\quad p=1,\ldots n\, , \quad k=1,\ldots, k_{p}\, .
\end{equation}
With this in mind, let us now consider an order-$(n+1)$ density built from a general linear combination of products of all these lower-order densities, \ie
\begin{align}
\tilde{\mathcal{R}}_{n+1}=\sum_{m=1}^{n}\sum_{k=1}^{k_m} \sum_{k'=1}^{k_{n+1-m}}C_{m,k,k'}\mathcal{R}^{(k)}_{m}\mathcal{R}^{(k')}_{n+1-m}\, ,
\end{align}
where we introduced the constants $C_{m,k,k'}$. 

We can ask now: is it possible to generate $n$ inequivalent GQTGs of order $(n+1)$ in this way? In order to answer this question, let us evaluate $\tilde{\mathcal{R}}_{n+1}$ on the single-function SSS ansatz and try to obtain all the possible on-shell GQTGs structures. The evaluation yields
\begin{align}
\tilde{\mathcal{R}}_{n+1}\Big|_{f}&=\sum_{m=1}^{n}\sum_{k=1}^{k_m} \sum_{k'=1}^{k_{n+1-m}}\sum_{i=0}^{m} \sum_{j=0}^{n+1-m}\alpha^{(k)}_{m,i} \alpha^{(k')}_{n+1-m,j}C_{m,k,k'} \mathcal{S}_{(m,i)}\mathcal{S}_{(n+1-m, j)}\\ \label{ramong}
&=\sum_{m=1}^{n}\sum_{i=0}^{m} \sum_{j=0}^{n+1-m}\tilde{C}_{m,i,j} \mathcal{S}_{(m,i)}\mathcal{S}_{(n+1-m, j)}\, ,
\end{align}
where we defined
\begin{equation}
\tilde{C}_{m,i,j}\equiv \sum_{k=1}^{k_m} \sum_{k'=1}^{k_{n+1-m}}\alpha^{(k)}_{m,i} \alpha^{(k')}_{n+1-m,j}C_{m,k,k'}\, .
\end{equation}
Now, since we are summing over all the $\alpha^{(k)}_{n,j}$ satisfying \eqref{alphaconstrv2} and $C_{m,k,k'}$ is an arbitrary tensor, note that this equality is equivalent to demanding that $\tilde{C}_{m,i,j}$ is an arbitrary tensor satisfying the following constraints
\begin{align}\label{Cconstr1}
\sum_{j=0}^{n+1-m}\tilde{C}_{m,i,j} [2(n+1-m) - D j ]&=0\, ,\quad \sum_{j=0}^{n+1-m} \tilde{C}_{m,i,j} j(j-1)=0\, ,\\
\label{Cconstr2}
\sum_{i=0}^{m}\tilde{C}_{m,i,j} (2m - Di)&=0\,,\quad \sum_{i=0}^{m} \tilde{C}_{m,i,j} i(i-1)=0\, .
\end{align}
In this way, we do not need to make reference to the  $\alpha^{(k)}_{n,j}$ anymore.  

Next, it is convenient to rearrange the sum in the following form, in terms of the index $l\equiv i+j$,
\begin{align}
\tilde{\mathcal{R}}_{n+1}\Big|_{f}&=\sum_{m=1}^{n}\sum_{l=0}^{n+1} \sum_{j=\max(l-m,0)}^{\min(l,n+1-m)}\tilde{C}_{m,l-j,j} \mathcal{S}_{(m,l-j)}\mathcal{S}_{(n+1-m, j)}\\
&=\sum_{l=0}^{n+1}\sum_{m=1}^{n} \sum_{j=0}^{n+1-m}\theta(l-j)\theta(j+m-l)\tilde{C}_{m,l-j,j} \mathcal{S}_{(m,l-j)}\mathcal{S}_{(n+1-m, j)}\, ,
\end{align}
where $\theta(x)\equiv 1$ if $x\ge0$ and $\theta(x)\equiv 0$ if $x<0$. Observe that the effect of the theta functions is to enforce that $i \geq 0$ and $i\leq m $, respectively, which in \req{ramong} is explicit from the $i$ sum.  Expanding the product $\mathcal{S}_{(m,l-j)}\mathcal{S}_{(n+1-m, j)}$ we get the following expression,
\begin{align}\notag
\tilde{\mathcal{R}}_{n+1}\Big|_{f}=&\sum_{l=0}^{n+1}\sum_{m=1}^{n} \sum_{j=0}^{n+1-m}\theta(l-j)\theta(j+m-l)\tilde{C}_{m,l-j,j} \\\notag
&  \times \bigg[\alpha_{l,m,j} B^{2+l}\psi ^{n-1-l}  +\beta_{l,m,j} B^{1+l} \psi ^{n-l}+\gamma_{l,m,j}B^l\psi ^{1-l+n}  
 \\
&\quad \, \, +\sigma_{l,m,j}r B' B^l \psi^{n-l} +\zeta_{l,m,j}rB'B^{l-1} \psi^{1-l+n} +\omega_{l,m,j}r^2\left(B'\right)^2B^{l-2}\psi ^{1-l+n} \bigg]\, ,
\end{align}
where
\begin{align}
\alpha_{l,m,j} \equiv &-4 (j-l+m) (-1+j+m-n)\, ,\\\notag
\beta_{l,m,j}  \equiv& -2 \big[1-4 j^2-5 l+4 m+D (-1+l-n)+4(l-m) (m-n)+n\\
&+4 j (1+l-2 m+n)\big]\, ,\\
\gamma_{l,m,j}  \equiv&+(-1+D-2 j+2 l-2 m) (-3+D+2 j+2m-2 n)\, ,\\
\sigma_{l,m,j}  \equiv&+2\left[2 j^2-j (1+2 l-2 m+n)+l (1-m+n)\right]\, ,\\
\zeta_{l,m,j}  \equiv&-\left[4 j^2-l (-3+D+2 m-2 n)-2 j (1+2 l-2 m+n)\right]\, ,\\
\omega_{l,m,j}  \equiv&- j (j-l)\, .
\end{align}

Finally, this can be recast as follows,
\begin{align}\notag
\tilde{\mathcal{R}}_{n+1}\Big|_{f}=\sum_{l=0}^{n+1} \left[\Gamma_{l}B^l\psi ^{1-l+n}+\Upsilon_{l}rB' B^{l-1}\psi ^{1-l+n}+\Omega_{l}r^2\left(B'\right)^2B^{l-2}\psi ^{1-l+n}\right]\, ,
\end{align}
where
\begin{align}\notag
\Gamma_{l}&\equiv \sum_{m=1}^{n}\sum_{j=0}^{n+1-m}\Bigg[\theta(l-2-j)\theta(j+m-l+2)\tilde{C}_{m,l-2-j,j} \alpha_{l-2,m,j}\\
&\quad +\theta(l-1-j)\theta(j+m-l+1)\tilde{C}_{m,l-1-j,j} \beta_{l-1,m,j} +\theta(l-j)\theta(j+m-l)\tilde{C}_{m,l-j,j} \gamma_{l,m,j} \Bigg]\, ,\\
\notag
\Upsilon_{l}&\equiv \sum_{m=1}^{n}\sum_{j=0}^{n+1-m}\Bigg[\theta(l-1-j)\theta(j+m-l+1)\tilde{C}_{m,l-1-j,j} \sigma_{l-1,m,j}\\
 & \quad +\theta(l-j)\theta(j+m-l)\tilde{C}_{m,l-j,j} \zeta_{l,m,j} \Bigg]\, ,\\ 
\Omega_{l}&\equiv \sum_{m=1}^{n}\sum_{j=0}^{n+1-m}\theta(l-j)\theta(j+m-l)\tilde{C}_{m,l-j,j} \omega_{l,m,j} \, .
\end{align}
Now, in order for this to be a GQTG we must have 
\begin{align}
\tilde{\mathcal{R}}_{n+1}\Big|_{f}&=\mathcal{S}^{(k)}_{n+1}=\sum_{l=0}^{n+1}\alpha^{(k)}_{n+1,l}  \mathcal{S}_{(n+1,l)}\\
&=\sum_{l=0}^{n+1}\alpha^{(k)}_{n+1,l} B^{l-1} \psi^{n-l} \left(l r \psi B'+B \psi (D+2 l-2 n-3)-2
   B^2 (l-n-1)\right)\\
   &=\sum_{l=0}^{n+1} \bigg[ B^{l} \psi^{n-l+1}\left(\alpha^{(k)}_{n+1,l}(D+2 l-2 n-3)-\alpha^{(k)}_{n+1,l-1}2(l-n-2)\right)\\
   &\quad \quad \quad +\alpha^{(k)}_{n+1,l}l r B'B^{l-1} \psi^{n-l+1} \bigg]\, ,
\end{align}
for some coefficients $\alpha^{(k)}_{n+1,l}$. Therefore, we have the equations
\begin{align}\label{rec1}
\Gamma_{l}=&\alpha^{(k)}_{n+1,l}(D+2 l-2 n-3)-\alpha^{(k)}_{n+1,l-1}2(l-n-2)\, , \\
\label{rec2}
\Upsilon_{l}=&l\alpha^{(k)}_{n+1,l}\, ,\\
\label{rec3}
\Omega_{l}=&0\, ,
\end{align}
for $l=0,\ldots , n+1$. In addition, the coefficients $\alpha^{(k)}_{n+1,l}$ should satisfy the constraints 
\begin{equation}\label{alphaconstrv3}
\sum_{l=0}^{n+1} \alpha^{(k)}_{n+1,l} (2n+2 - Dl)=0\, ,\quad \quad \sum_{l=0}^{n+1} \alpha^{(k)}_{n+1,l} l(l-1)=0\, ,
\end{equation}
but note that these must arise as consistency conditions in order for the system of equations to have solutions.
Then, the question is whether the system of equations for the tensor $\tilde C_{m,i,j}$ given by Eqs.~\eqref{Cconstr1},  \eqref{Cconstr2}, \eqref{rec1}, \eqref{rec2}, \eqref{rec3} has solutions for \emph{any} value of the $\alpha^{(k)}_{n+1,l}$ satisfying the constraints \eqref{alphaconstrv3}. If that is the case, then we have proven the existence of $n$ different GQTGs at order $n+1$ which, as we saw earlier, is the maximum possible number of GQTGs at that order. 

The number of equations to be solved for fixed $n$ ---namely, the number of equations required for establishing the existence of $n$ densities of order $(n+1)$--- and the number of unknowns ($\tilde C_{m,i,j}$)  read, respectively
\begin{equation}
\# \text{ equations} = \frac{12+n(11+3n)}{2}\, , \quad \# \text{ unknowns} =\frac{n(n+2)(n+7)}{6}\, .
\end{equation}
The former is greater than the latter as long as $n<5.10421$ and smaller for greater values of $n$. Observe that while the number of equations grows as $\sim n^2$, the number of unknowns grows as $\sim n^3$. Here, the number of unknowns is the number of constants available to be fixed in order for the GQTG conditions to be satisfied, and so having more unknowns than equations means that we have more than enough freedom to impose all the conditions. Hence, as long as we are able to show that the $(n-1)$ different classes of GQTG exist for $n\leq 6$ using other methods, this result shows that they will generally exist for $n > 6$.  In practice, solving this system of equations explicitly for any $n\geq 6$ and $D$ is challenging, 
Nevertheless, the resolution for explicit values of $n$ and $D$ is straightforward with the help of a computer algebra system. Doing this, we have checked that there is a solution for any consistent value of the $\alpha^{(k)}_{n+1,l}$ in any $D$ as long as $n\ge 6$.\footnote{In practice, we have checked this explicitly for $n=6$ and general $D$ and for $n=7,\dots, 20$ in $D=5,6,7$. } 

In sum, our results here imply that, if $(n-1)$ inequivalent GQTGs exist for $n=1,\ldots, 6$, then, $(n-1)$ inequivalent densities will exist for every order $n\geq 6$. In Appendix \ref{densiii} we have provided explicit examples of all the inequivalent classes of GQTGs up to $n=6$ for $D=5,6$, so  this proves that there are $(n-1)$ inequivalent GQTGs at every order $n\ge 2$ in those cases. The construction of explicit $n\leq 6$ densities of all the different classes for other values of $D$ can be analogously performed (although it requires some non-trivial computational effort in each case) so we are highly confident that our results apply for general $D\geq 7$ as well.

On the other hand, note that our argument here does not work in $D=4$. Indeed, we have found no evidence for the existence of additional inequivalent GQTGs (besides the one known prior to this paper \cite{Hennigar:2016gkm,PabloPablo2,PabloPablo4,Bueno:2019ycr}) up to order $6$ in that case. This strongly suggests that in $D=4$ there is a single type of GQTG at every curvature order although a rigorous proof of this fact would require some additional work.

\section{Black Hole Thermodynamics}\label{GQTGthermo}
In this section we study thermodynamic aspects of GQTGs in an as general as possible fashion. First we show that the first law of black hole mechanics is satisfied by the black hole solutions of general GQTGs. Then, we will show that thermodynamic magnitudes of at least one class of genuine GQTGs can be, similarly to the Lovelock and quasi-topological cases, expressed in terms of the characteristic polynomial which embeds maximally symmetric backgrounds in the theory and the on-shell Lagrangian. 

\subsection{The first law for general GQTGs}

Here we wish to understand the first law of thermodynamics for all possible GQTGs. We will begin by working directly with Eq.~\eqref{fnj}, without imposing the constraints on the couplings given in Eqs.~\eqref{cond1} and \eqref{cond2} at this time.
The integrated field equations of the putative theory can be written in the form
\be 
\sum_{n=0}^{n_{\rm max}} \sum_{j=0}^{n} \alpha_{n,j} \mathcal{F}_{(n,j)} = - \frac{8 \pi  G M}{\Omega_{D-2}} \, ,
\ee
where the parameter $M$ is the black hole mass \cite{Arnowitt:1960es,Arnowitt:1960zzc,Arnowitt:1961zz,Deser:2002jk}. At a black hole horizon, where $f(\rp) = 0$, the above equation can be expanded to yield the following constraints:
\begin{align}
M &= \frac{\Omega_{D-2}}{16 \pi G} \sum_{n=0}^{n_{\rm max}} \sum_{j=0}^{n} \alpha_{n,j} (j-1) k^{n-j} \rp^{D-2n - 1} (-2 \pi \rp T)^j \, ,
\\
0 &= \sum_{n=0}^{n_{\rm max}} \sum_{j=0}^{n} \alpha_{n,j} (D-2n +j-1) k^{n-j} (-2 \pi \rp T)^j \rp^{D-2n-2} \, .
\end{align}
where the temperature satisfies $T=f'(r_+)/(4\pi)$.
The first equation gives the black hole mass in terms of the temperature $T$ and the horizon radius $\rp$, while the second provides a relationship between $T$ and $\rp$. 

The other ingredient we need is the black hole entropy. This should be computed according to Wald's formula \cite{Wald:1993nt,Iyer:1994ys}
\be 
S = -2\pi \int_{\mathcal{H}} \diff^{D-2}  x  \sqrt{h} \, P_{ab}{}^{cd} \varepsilon^{ab}\varepsilon_{cd} \, ,
\ee
where $\varepsilon_{ab}$ is the binormal to the horizon $\mathcal{H}$. 
Using the technology introduced in~\cite{Bueno:2019ycr}, this can be computed without knowledge of the covariant form of the Lagrangian. The key insight is that the tensor $P_{ab}{}^{cd}$ can be computed from the on-shell Lagrangian and must take the form
\be \label{Pabcd}
\left.\tensor{P}{_{cd}^{ab}}\right|_{f} = P_1 T^{[a}_{[c} T^{b]}_{d]} +P_2   T^{[a}_{[c} \sigma^{b]}_{d]} +P_3 \sigma^{[a}_{[c} \sigma^{b]}_{d]} \, ,
\ee
where
\begin{equation}
P_1\equiv  -  \frac{\partial {\cal R}_{(n)}|_f}{\partial f''} \, ,\quad P_2 \equiv - \frac{r}{D-2} \frac{\partial {\cal R}_{(n)}|_f}{\partial f'}  \, , \quad P_3\equiv - \frac{ r^2 }{(D-2)(D-3)} \frac{\partial {\cal R}_{(n)}|_f}{\partial f }\, .
\end{equation}
For the case of the static and spherically symmetric black holes considered here, the horizon binormal is given by $\varepsilon_{ab} = 
 2 r_{[a} t_{b]}$ with $r^a$ and $t^b$ the unit spacelike and timelike normal vectors. A calculation then gives
\be 
S = -4 \pi \Omega_{D-2} \rp^{D-2} \left[\frac{\partial \mathcal{L}}{\partial f''} \right]_{r = \rp} = \frac{\Omega_{D-2}}{8 G} \sum_{n=0}^{n_{\rm max}} \sum_{j=0}^{n} \alpha_{n,j} j k^{n-j} (-2 \pi \rp T)^{j-1} \rp^{D-2n} \, .
\ee
It is then straight-forward to show that the first law of thermodynamics
\be 
\diff M = T \diff S
\ee
holds independent of any conditions placed on the couplings $\alpha_{n,j}$. This fact is somewhat surprising because, as discussed earlier, it is only when certain constraints are obeyed by the couplings that a genuine, covariant construction for the Lagrangian can be built based on curvature invariants. However, these same constraints are unnecessary to obtain a valid first law. 

Despite the fact that the coupling constraints are not necessary to obtain a valid first law, it is still possible to understand them from a thermodynamic perspective. For this, the natural starting point is the free energy, which reads
\be 
F = \frac{\Omega_{D-2}}{16 \pi G} \sum_{n, j} \alpha_{n, j} k^{n-j}(-2 \pi T)^j \rp^{D-1 -2n+j} \, .
\ee
From the free energy, the equation that relates the temperature and horizon radius can be obtained according to
\be 
\frac{\partial F}{\partial \rp} = 0 \, ,
\ee
while the mass and entropy can then be verified to follow in the usual way. The constraints on the couplings enforce the following conditions on the free energy:
\begin{eqnarray}
	F-T\frac{\partial F}{\partial T}-\frac{\rp}{D-1}\frac{\partial F}{\partial \rp}\Bigg|_{2\pi T \rp=-k}&=&0\, ,\\
	\frac{\partial^2 F}{\partial T^2}\Bigg|_{2\pi T \rp =-k}&=&0\, ,
\end{eqnarray}
where it is to be noted that the derivatives here are to be computed without assuming any relationship between $\rp$ and $T$.

These expressions above, phrasing the coupling constraints in as properties of the free energy, can be reinterpreted as statements about massless hyperbolic black holes. The static black hole with metric function
\be 
f(r)=-1 + \frac{r^2}{L_{\star}^2} 
\ee
is pure AdS space in a particular slicing. In terms of the parameters we have been using, this corresponds to $k=-1$, $\rp = L_\star$ and $T = 1/(2 \pi L_\star)$, therefore satisfying the condition $2 \pi T \rp = - k$. In this language, as we will see explicitly below, the first of the two constraints on the free energy actually ensures that the mass of this black hole vanishes. The second constraint on the free energy does not have as direct of an interpretation in terms of the thermodynamic properties of this black hole, but one could imagine it is a statement about fluctuations.

\subsection{A unified picture of the thermodynamics?}

Lovelock and quasi-topological gravities are, by comparison to alternatives, rather simple extensions of general relativity, especially in the context of static, spherically symmetric black holes. Within our parameterization, the coupling constants $\alpha_{n,j}$ to achieve the on-shell Lagrangian for Lovelock and quasi-topological theories amounts to the choice~\eqref{qt}. For these theories, as has long been known in the case of Lovelock \cite{Boulware:1985wk,Wheeler:1985nh,Wheeler:1985qd}, the field equations for a static, spherically symmetric black hole take the form
\be 
M = \frac{(D-2)\Omega_{D-2} r^{D-1}}{16 \pi G L^2} h(y) \, , \quad y \equiv \frac{(f(r) -k) L^2}{r^2} \, .
\ee
The function $h(x)$ appearing here is the same function that determines the vacua of the theory, \ie the field equations for the maximally symmetric solutions of the theory. This ``embedding function'' or ``characteristic polynomial'' is related to the Lagrangian of the theory evaluated on a maximally symmetric background \cite{Aspects,Bueno:2018yzo}
\be\label{embed} 
h(x) = \frac{16 \pi G L^2}{(D-1)(D-2)} \left[\mathcal{L}(x) - \frac{2}{D} x \mathcal{L}'(x) \right]\, ,
\ee
where here $x$ is related to the curvature of the maximally symmetric background according to
\be\label{mssCurv} 
R_{ab}{}^{cd} = - \frac{2 x}{L^2} \delta_{[a}^c \delta_{b]}^d \, ,
\ee
and $\mathcal{L}(x)$ corresponds to the Lagrangian of the theory evaluated for the curvature~\eqref{mssCurv}.

The fact that the field equations can be written in terms of the embedding function naturally leads to some simple and universal  expressions for black hole thermodynamics:
\begin{align}
M &= \frac{(D-2)\Omega_{D-2} \rp^{D-1}}{16 \pi G L^2} h(\yp) \, , \quad \yp \equiv -\frac{kL^2}{\rp^2} \, , 
\nonumber\\
S &= - \frac{4 \pi \Omega_{D-2} L^2 \rp^{D-2}}{D(D-1) } \mathcal{L}'(\yp) \, .
\end{align}
These relationships are expressed here in their simplest possible forms, but of course can be massaged using the identity~\eqref{embed} and its derivatives, along with the constraint 
\be 
0 = (D-1)k h(\yp) - 2 \yp (2 \pi \rp T + k) h'(\yp) \, ,
\ee
which can be used to isolate for the temperature, if desired. 

It is natural to wonder whether similar relationships hold for the more complicated generalized quasi-topological theories, or whether this result for Lovelock and quasi-topological theories was an artefact of their simplicity. Here we will provide evidence that this is indeed possible, though the situation is more involved than the Lovelock and quasi-topological cases.

Consider the family of theories identified according to the following choices of couplings:
\begin{align}
\alpha_{n, n-j} = \frac{(D-4)^{j-1} n! \left[\left(n - j - 2 \right) D - 4(n-2) \right]}{2^{2j} \,  j! \, (n-j)! \, (n-2)} \alpha_{n,n} \, .
\end{align}
In general dimensions, this corresponds to the family of theories for which an explicit covariant formulation was identified in~\cite{Bueno:2019ycr}. These couplings satisfy the necessary constraints~\eqref{cond1} and~\eqref{cond2}, and in addition define a family of GQTG theories for which the free energy can be written as,
\be 
F = - \frac{(D-2) \Omega_{D-2} \rp^{D-1}}{16 \pi G L^2} h(x_+) - \frac{4 L^2 \rp^{D-3}}{D^2(D-1)} \left[(D-2)k + (D-4) \pi \rp T \right] \mathcal{L}'(x_+)
\ee
where
\be 
x_+  \equiv \frac{8 \pi T L^2}{\rp D} - \frac{(D-4) k L^2}{\rp^2 D} \, .
\ee
From this form of the free energy, the full thermodynamic properties for this class of theories can be derived. We obtain for the mass and relationship between the temperature and horizon radius the following two results:
\begin{align}
M =& \, \frac{(D-2) \Omega_{D-2} \rp^{D-1}}{16 \pi G L^2} h(x_+) - \frac{(D-2) \Omega_{D-2} \rp^{D-3}}{4 \pi G D} [2 \pi \rp T + k] h'(x_+) 
\nonumber\\
&+ \frac{(D-4) \Omega_{D-2} L^4 \rp^{D-5}}{D^3(D-1)} [2 \pi \rp T + k]^2 \mathcal{L}''(x_+) \, , 
\\
0 =&\, (D-1)(D-2) h(x_+) - \frac{2 (D-2)^2 L^2}{\rp^2 D} [2 \pi \rp T + k] h'(x_+) 
\nonumber\\
&- \frac{8 (D-4) L^6}{D^3 (D-1) \rp^4} [2\pi \rp T + k]^2 \left( 16 \pi G \mathcal{L}''(x_+)\right)  \, ,
\end{align}
while the entropy can be simply obtained from the above according to $S = (M-F)/T$.

It is a bit interesting that the thermodynamic properties of black holes can be encoded in terms of the embedding function $h(x)$ and the Lagrangian of the theory $\mathcal{L}(x)$ evaluated on an auxiliary maximally symmetric vacuum spacetime with curvature given by $x_+/L^2$. There is one case where this result is somewhat natural, and this is the case of massless hyperbolic black holes where $f = -1 + r^2/L_{\star}^2$. Of course, this choice of metric function amounts to a pure AdS space in a particular slicing. One has $k=-1$, $T = 1/(2 \pi L_{\star})$, and $x_+ = L^2/L_{\star}^2$. In this case, the only non-trivial field equation demands that $h(x_+) = 0$, which in turn demands that $M = 0$. 

Next, note that considerable simplification occurs in $D = 4$. In this case, the situation reduces to that first studied in~\cite{PabloPablo4}. In that case, the couplings are given by
\be 
\alpha_{n, n-1} = -\frac{n }{n-2} \alpha_{n, n} \, , \quad \alpha_{2, j} = 0 \quad \forall j \, , \quad \text{and} \quad \alpha_{n, j} = 0 \quad \forall j \neq n, n-1 \, , \forall n \ge 3 \, .
\ee 
The thermodynamic relations in this case simplify to
\begin{align}
M &= \frac{\Omega_{D-2} \rp^3}{8 \pi G L^2} h(\xp) - \frac{\Omega_{D-2} \rp}{8 \pi G} \left[2 \pi \rp T + k \right] h'(\xp) \, , 
\\
S &= \frac{\Omega_{D-2} k \rp L^2}{6 T} \mathcal{L}'(\xp) - \frac{\Omega_{D-2} \rp}{8 \pi G T} \left[2 \pi \rp T + k \right] h'(\xp) \, , 
\end{align}
and the constraint that determines the temperature in terms of the horizon radius reads
\be 
0 = \frac{-3 \rp^2}{L^2} h(\xp) + \left[2 \pi \rp T + k \right] h'(\xp) \, .
\ee

It seems likely that the thermodynamics of each family of GQTG can be obtained in this way, though we will leave that full analysis for future work. Nonetheless, we can make a few general remarks, based on the connection with massless hyperbolic black holes. For any given family of GQTGs, the mass must have a term proportional to $h(x)$ followed by a series of terms with powers that vanish for the massless hyperbolic black hole. For example, the simplest possibility would be $(2 \pi \rp T + k)$ raised to various powers, multiplying derivatives of $h$ and $\mathcal{L}$. Similarly, the entropy must have a term proportional to $\mathcal{L}'(x)$, followed by a series of terms that vanish for the massless hyperbolic black hole, just as above. Lastly, the argument $x$ must be a function of $\rp$, $T$ and $k$ that limits to $L^2/L_{\star}^2$ for the massless hyperbolic black hole. For example, allowing for a linear dependence on the parameters, the most general option is the one-parameter family 
\be 
x_+ = \frac{2 \pi T L^2 \beta}{\rp} + \frac{(\beta-1) k L^2}{\rp^2} \, .
\ee
This linear relationship recovers the result for Lovelock/quasi-topological gravity (with $\beta = 0$) and the GQTG family we have presented above (with $\beta = 4/D$). Preliminary calculations have suggested that other GQTG families may require a more complicated dependence than this.

\section{Final comments} \label{discuss}

In this work, we have completed the structural analysis of generalized quasi-topological gravities, proving that at order $n$ in curvature there exist $n-1$ distinct GQTGs provided $D > 4$. In the case of $D = 4$, our results strongly suggest that there is a single  (unique up to addition of trivial densities) GQTG family corresponding to that identified in~\cite{PabloPablo4}. To achieve this, we first derived an upper bound, based on the fact that an on-shell GQTG density must be a polynomial in the three independent terms appearing in the Riemann curvature for a static, spherically symmetric background. This upper bound, which holds independent of any knowledge of the covariant form of the densities, was then refined by demanding of the putative theories additional properties that must hold for a true covariant density. Finally, we proved the refined estimate to be exact using arguments based on recurrence formulas, like those introduced in~\cite{Bueno:2019ycr}. In order for our argument to hold, it is required that $n-1$ densities exist for $n=2,3,4,5,6$, which then implies existence for all $n>6$. Such $n-1$ densities for the lowest curvature orders can be constructed explicitly for $D\geq 5$ but not for $D=4$, in which case we have verified that there is always a unique density for every $n=2,\dots,6$. The argument for higher $n$ then fails for $D=4$. While it could in principle be possible that additional inequivalent densities exist in $D=4$ for higher orders ---and our construction involving products of lower-order densities was not general enough to capture them--- we find this possibility highly unlikely.

In addition, we have provided a basic analysis of the thermodynamic properties of black holes in all possible theories, confirming that the first law is satisfied. Perhaps the most interesting result in this direction is the strong evidence that the thermodynamics of black holes in any GQTG may be expressible in terms of the same function that determines the vacua of the theory, just like in Lovelock and quasi-topological gravities. Why the thermodynamics of black holes in these theories is encoded in the curvature of some axillary maximally symmetric space remains mysterious to us, and may be worth further investigation. More pragmatically, such closed-form and universal expressions provide a simple means by which the thermodynamics could be studied when an infinite number of higher-curvature corrections are simultaneously included. 

As a by-product, our work has identified $(n-2)$ hitherto unknown families of GQTGs in $D > 4$. Going forward, it would be interesting to understand how the properties of black hole solutions differ between these different families, or whether there exist universal features, such as occurs in $D = 4$~\cite{PabloPablo4}. Moreover, the methods we have used to upper bound the number of distinct theories may generalize to allow for a similar analysis to be carried out when there is non-minimal coupling between gravity and matter fields.

\section*{Acknowledgments}
 In some cases, calculations performed in the manuscript have been facilitated by Maple and Mathematica, utilizing the specialized packages GRTensor and xAct~\cite{xact}. The work of PB was partially supported by the Simons foundation through the It From Qubit Simons collaboration.  The work of PAC is supported by a postdoctoral fellowship from the Research
Foundation - Flanders (FWO grant 12ZH121N).  The work of RAH is supported physically by planet Earth through the electromagnetic and gravitational interactions, and received the support of a fellowship from ``la Caixa” Foundation (ID 100010434) and from the European Union’s Horizon 2020 research and innovation programme under the Marie Skłodowska-Curie grant agreement No 847648” under fellowship code LCF/BQ/PI21/11830027. The work of JM is funded by the Agencia Nacional de Investigaci\'on y Desarrollo (ANID) scholarship No. 21190234 and by Pontificia Universidad Cat\'olica de Valpara\'iso.
 
\appendix

\section{Explicit covariant densities for $n=4,5,6$ in $D=5,6$} \label{densiii}
In this appendix we present explicit GQT covariant densities of each of the $(n-2)$ existing types for $n=4,5,6$ in $D=5$ and $D=6$. 

At quartic order, examples of representatives of the two inequivalent classes of GQT densities in $D=5$ are (we use Roman numbers to label the different types)
\begin{align}
\mathcal{S}_{[D=5,n=4]}^{\rm  I}=&+12 R\indices{^a^b^c^d} R\indices{_a_b^e^f} R\indices{_c^g_e^h} R\indices{_d_g_f_h}+3 R\indices{^a^b^c^d} R\indices{_a^e_c^f} R\indices{_b_g_d_h}
   R\indices{_e^g_f^h} \nonumber\\
   & -6 R\indices{^a^b^c^d} R\indices{_a^e_c^f} R\indices{_e^g_b^h} R\indices{_f_g_d_h}-9 R\indices{^a^b} R\indices{_c^h_e_a}
   R\indices{_d_h_f_b} R\indices{^c^d^e^f}+R R\indices{_a^c_b^d} R\indices{_e^a_f^b} R\indices{_c^e_d^f} \, , \\\
   \mathcal{S}_{[D=5,n=4]}^{\rm  II}=&+4 R\indices{^a^b^c^d} R\indices{_a_b^e^f} R\indices{_c_e^g^h} R\indices{_d_f_g_h}+30 R\indices{^a^b^c^d} R\indices{_a_b^e^f} R\indices{_c^g_e^h}R\indices{_d_g_f_h} \nonumber\\
& -11 R\indices{^a^b^c^d} R\indices{_a^e_c^f} R\indices{_e^g_b^h} R\indices{_f_g_d_h}-16 R\indices{^a^b} R\indices{_c^h_e_a}R\indices{_d_h_f_b} R\indices{^c^d^e^f}-R\indices{^a^b} R\indices{_c_d^h_a} R\indices{_e_f_h_b} R\indices{^c^d^e^f} \nonumber\\
&-3 R\indices{^a^b} R\indices{_a^c_b^d}R\indices{_e_f_h_c} R\indices{^e^f^h_d}+3 R\indices{^a^b} R\indices{^c^d} R\indices{^e_a^f_b} R\indices{_e_c_f_d}
+R\indices{^a^b} R\indices{^c^d} R\indices{^e_a^f_c}R\indices{_e_b_f_d}\, ,
\end{align}
which evaluated on the single-function ansatz reduce to linear combinations of $\mathcal{S}_{(4,j)}|_f$, as defined in \req{rj}, with
\begin{align}
\tau_{[D=5,n=4]}^{\rm  I}&=4\tau_{(4,1)}+12 \tau_{(4,3)}-6 \tau_{(4,4)} \, , \\  \tau_{[D=5,n=4]}^{\rm  II}&=6 \tau_{ (4,2)}-\tau_{(4,4)}\, ,
\end{align}
respectively. It is straightforward to check that both satisfy conditions \req{cond1} and \req{cond2}. In $D=6$, we find
\begin{align}
\mathcal{S}_{[D=6,n=4]}^{\rm  I}=&+15 R\indices{^a^b^c^d} R\indices{_a_b^e^f} R\indices{_c_e^g^h} R\indices{_d_f_g_h}+20 R\indices{^a^b^c^d} R\indices{_a_b^e^f} R\indices{_c^g_e^h}
   R\indices{_d_g_f_h}-4 R\indices{^a^b^c^d} R\indices{_a^e_c^f} R\indices{_b_g_d_h} R\indices{_e^g_f^h}\nonumber\\
   &-36 R\indices{^a^b^c^d} R\indices{_a^e_c^f}
   R\indices{_e^g_b^h} R\indices{_f_g_d_h}+48 R\indices{^a^b} R\indices{_c^h_e_a} R\indices{_d_h_f_b} R\indices{^c^d^e^f}-8 R\indices{^a^b} R\indices{_c_d^h_a}
   R\indices{_e_f_h_b} R\indices{^c^d^e^f}\nonumber\\
   &-8 R R\indices{_a^c_b^d} R\indices{_e^a_f^b} R\indices{_c^e_d^f}+8 R\indices{^a^b} R\indices{^c^d} R\indices{^e_a^f_c}
   R\indices{_e_b_f_d} \, , \\
   \mathcal{S}_{[D=6,n=4]}^{\rm  II}=&-5 R\indices{^a^b^c^d} R\indices{_a_b^e^f} R\indices{_c_e^g^h} R\indices{_d_f_g_h}-28 R\indices{^a^b^c^d} R\indices{_a_b^e^f} R\indices{_c^g_e^h}
   R\indices{_d_g_f_h}-20 R\indices{^a^b^c^d} R\indices{_a^e_c^f} R\indices{_b_g_d_h} R\indices{_e^g_f^h}\nonumber\\
   &+52 R\indices{^a^b^c^d} R\indices{_a^e_c^f}
   R\indices{_e^g_b^h} R\indices{_f_g_d_h}-16 R\indices{^a^b} R\indices{_c^h_e_a} R\indices{_d_h_f_b} R\indices{^c^d^e^f}+8 R\indices{^a^b} R\indices{_c_d^h_a}
   R\indices{_e_f_h_b} R\indices{^c^d^e^f}\nonumber\\
   &-8 R\indices{^a^b} R\indices{_a^c_b^d} R\indices{_e_f_h_c} R\indices{^e^f^h_d}+8 R\indices{^a^b} R\indices{^c^d}
   R\indices{^e_a^f_b} R\indices{_e_c_f_d}-8 R\indices{^a^b} R\indices{^c^d} R\indices{^e_a^f_c} R\indices{_e_b_f_d} \, ,
\end{align}
and for those
\begin{align}
\tau_{[D=6,n=4]}^{\rm  I}&=\tau_{(4, 4)} - 2 \tau_{(4, 3)} - 2\tau_{(4, 1)} \, .\\
  \tau_{[D=6,n=4]}^{\rm  II}&=\tau_{(4, 4)} - 4 \tau_{(4, 3)} - 6\tau_{(4, 2)} .
\end{align}

 At quintic order, examples of the three inequivalent classes read
\begin{align}
\mathcal{S}_{[D=5,n=5]}^{\rm  I}=&+
3235 R^5 - 28409 R^3 R_{a}{}^{b} R_{b}{}^{a} + 46980 R^2 R_{a}{}^{c}R_{b}{}^{a} R_{c}{}^{b} - 93522 R R_{a}{}^{d} R_{b}{}^{a} R_{c}{}^{b} R_{d}{}^{c}\nonumber\\
&+ 11928 R_{a}{}^{b} R_{b}{}^{a} R_{c}{}^{e} R_{d}{}^{c} R_{e}{}^{d} + 98700 R R_{b}{}^{a} R_{d}{}^{b} R_{e}{}^{c} 
R_{ac}{}^{de} + 2870 R^3 R_{ab}{}^{cd} R_{cd}{}^{ab}\nonumber\\
&+ 52080 
R_{a}{}^{b} R_{b}{}^{a} R_{e}{}^{c} R_{f}{}^{d} R_{cd}{}^{ef} - 
151200 R R_{c}{}^{a} R_{d}{}^{b} R_{ab}{}^{ef} R_{ef}{}^{cd}\nonumber\\
& + 137655 
R R_{b}{}^{a} R_{c}{}^{b} R_{ad}{}^{ef} R_{ef}{}^{cd}- 5845 R 
R_{a}{}^{b} R_{b}{}^{a} R_{cd}{}^{ef} R_{ef}{}^{cd} \nonumber\\
&- 23940 
R_{a}{}^{b} R_{b}{}^{a} R_{d}{}^{c} R_{ce}{}^{fg} R_{fg}{}^{de} \, , \\
\mathcal{S}_{[D=5,n=5]}^{\rm  II}=&+10505 R^5 - 98197 R^3 R_{a}{}^{b} R_{b}{}^{a} + 242460 R^2 
R_{a}{}^{c} R_{b}{}^{a} R_{c}{}^{b} - 362526 R R_{a}{}^{d} 
R_{b}{}^{a} R_{c}{}^{b} R_{d}{}^{c}\nonumber\\ 
&+ 77784 R_{a}{}^{b} R_{b}{}^{a} 
R_{c}{}^{e} R_{d}{}^{c} R_{e}{}^{d} + 77700 R R_{b}{}^{a} R_{d}{}^{b} 
R_{e}{}^{c} R_{ac}{}^{de} + 1120 R^3 R_{ab}{}^{cd} R_{cd}{}^{ab}\nonumber\\
&+ 
139440 R_{a}{}^{b} R_{b}{}^{a} R_{e}{}^{c} R_{f}{}^{d} R_{cd}{}^{ef} 
- 173880 R R_{c}{}^{a} R_{d}{}^{b} R_{ab}{}^{ef} R_{ef}{}^{cd} \nonumber\\
& + 
194985 R R_{b}{}^{a} R_{c}{}^{b} R_{ad}{}^{ef} R_{ef}{}^{cd}+ 12355 
R R_{a}{}^{b} R_{b}{}^{a} R_{cd}{}^{ef} R_{ef}{}^{cd} \nonumber\\
&- 104580 
R_{a}{}^{b} R_{b}{}^{a} R_{d}{}^{c} R_{ce}{}^{fg} R_{fg}{}^{de} - 
15120 R R_{b}{}^{a} R_{ad}{}^{bc} R_{ce}{}^{fg} R_{fg}{}^{de}\nonumber\\
&- 3780 
R R_{b}{}^{a} R_{ac}{}^{fg} R_{de}{}^{bc} R_{fg}{}^{de} + 11340 
R_{a}{}^{b} R_{b}{}^{a} R_{cd}{}^{gh} R_{ef}{}^{cd} R_{gh}{}^{ef} \, , \\
\mathcal{S}_{[D=5,n=5]}^{\rm  III}=&
-108751900 R^5 + 1026499979 R^3 R_{a}{}^{b} R_{b}{}^{a} - 2724816480 
R^2 R_{a}{}^{c} R_{b}{}^{a} R_{c}{}^{b}\nonumber\\ &+ 3743976918 R R_{a}{}^{d} 
R_{b}{}^{a} R_{c}{}^{b} R_{d}{}^{c} - 981715812 R_{a}{}^{b} 
R_{b}{}^{a} R_{c}{}^{e} R_{d}{}^{c} R_{e}{}^{d} \nonumber\\ &+ 241948812 R 
R_{b}{}^{a} R_{d}{}^{b} R_{e}{}^{c} R_{ac}{}^{de}+ 11124379 R^3 
R_{ab}{}^{cd} R_{cd}{}^{ab} \nonumber\\ &+ 2523150 R R_{ab}{}^{cd}{}^2 
R_{cd}{}^{ab}{}^2 - 1472417016 R_{a}{}^{b} R_{b}{}^{a} R_{e}{}^{c} 
R_{f}{}^{d} R_{cd}{}^{ef}\nonumber\\
& + 442592640 R R_{c}{}^{a} R_{d}{}^{b} 
R_{ab}{}^{ef} R_{ef}{}^{cd} - 1009017009 R R_{b}{}^{a} R_{c}{}^{b} 
R_{ad}{}^{ef} R_{ef}{}^{cd}\nonumber\\
& - 199666439 R R_{a}{}^{b} R_{b}{}^{a} 
R_{cd}{}^{ef} R_{ef}{}^{cd} + 1327705722 R_{a}{}^{b} R_{b}{}^{a} 
R_{d}{}^{c} R_{ce}{}^{fg} R_{fg}{}^{de}\nonumber\\
&- 7998480 R R_{b}{}^{a} 
R_{ad}{}^{bc} R_{ce}{}^{fg} R_{fg}{}^{de} + 151439400 R R_{b}{}^{a} 
R_{ac}{}^{fg} R_{de}{}^{bc} R_{fg}{}^{de}\nonumber\\
& - 197676360 R_{a}{}^{b} 
R_{b}{}^{a} R_{cd}{}^{gh} R_{ef}{}^{cd} R_{gh}{}^{ef} +35700000 
R_{ab}{}^{cd} R_{cd}{}^{ab} R_{ej}{}^{gh} R_{fh}{}^{ij} R_{gi}{}^{ef} \nonumber\\&
+ 121836960 R_{b}{}^{a} R_{ad}{}^{bc} R_{cf}{}^{de} R_{eg}{}^{hi} 
R_{hi}{}^{fg} - 89250 R_{ab}{}^{cd} R_{cd}{}^{ab} R_{ef}{}^{ij} 
R_{gh}{}^{ef} R_{ij}{}^{gh}\, ,
\end{align}
And for them
\begin{align}
\tau_{[D=5,n=5]}^{\rm  I}&=+2 \tau_{(5,0)}-\tau_{(5,1)}-12\tau_{(5,2)}-10 \tau_{(5,3)}+2 \tau_{(5,4)}+3 \tau_{(5,5)}\, , \\
\tau_{[D=5,n=5]}^{\rm  II}&=-5 \tau_{(5,0)}+4 \tau_{(5,1)}+18 \tau_{(5,2)}+4 \tau_{(5,3)}-5 \tau_{(5,4)}\, , \\
\tau_{[D=5,n=5]}^{\rm  III}&=+45\tau_{(5,0)}-46 \tau_{(5,1)}+44 (\tau_{(5,3)}-3 \tau _{(5,2)}\, .
\end{align}
For $D=6$, we find
\begin{align}
\mathcal{S}_{[D=6,n=5]}^{\rm  I}=&-123946191482880 R_{a}{}^{b} R_{b}{}^{a} R_{c}{}^{e} R_{d}{}^{c} \
R_{e}{}^{d} + 1472406237369312 R_{a}{}^{d} R_{b}{}^{a} R_{c}{}^{b} \
R_{d}{}^{c} R\nonumber\\ &- 1080277675306560 R_{a}{}^{c} R_{b}{}^{a} R_{c}{}^{b} \
R^2 + 162174148310040 R_{a}{}^{b} R_{b}{}^{a} R^3\nonumber\\
&- 11444059832562 \
R^5 + 1702982503075584 R_{b}{}^{a} R_{d}{}^{b} R_{e}{}^{c} R \
R_{ac}{}^{de}\nonumber\\
&+ 75220642409760 R^3 R_{ab}{}^{cd} R_{cd}{}^{ab} + \
12994390356246 R \left(R_{ab}{}^{cd}{} R_{cd}{}^{ab}{}\right)^2\nonumber\\
&- 941724825600 \
R_{a}{}^{b} R_{b}{}^{a} R_{e}{}^{c} R_{f}{}^{d} R_{cd}{}^{ef} - \
1826681030324352 R_{c}{}^{a} R_{d}{}^{b} R R_{ab}{}^{ef} \
R_{ef}{}^{cd}\nonumber\\
&+ 1161324617394816 R_{b}{}^{a} R_{c}{}^{b} R \
R_{ad}{}^{ef} R_{ef}{}^{cd} - 402058236112056 R_{a}{}^{b} R_{b}{}^{a} \
R R_{cd}{}^{ef} R_{ef}{}^{cd}\nonumber\\
&+ 796036321619712 R_{b}{}^{a} R \
R_{ad}{}^{bc} R_{ce}{}^{fg} R_{fg}{}^{de}\nonumber\\
&- 226245709813248 \
R_{b}{}^{a} R R_{ac}{}^{fg} R_{de}{}^{bc} R_{fg}{}^{de}\nonumber\\
&- \
2713887813611520 R_{ag}{}^{cd} R_{bi}{}^{ef} R_{ce}{}^{ab} \
R_{dj}{}^{gh} R_{fh}{}^{ij}\nonumber\\
&+ 5441837051289600 R_{ag}{}^{cd} \
R_{bi}{}^{ef} R_{ce}{}^{ab} R_{dh}{}^{ij} R_{fj}{}^{gh}\nonumber\\
&- \
8516393811394560 R_{ag}{}^{cd} R_{bh}{}^{ij} R_{ce}{}^{ab} \
R_{di}{}^{ef} R_{fj}{}^{gh}\nonumber\\
&- 9075154990067712 R_{aj}{}^{gh} \
R_{bd}{}^{ij} R_{ce}{}^{ab} R_{fg}{}^{cd} R_{hi}{}^{ef}\, ,\\
\mathcal{S}_{[D=6,n=5]}^{\rm  II}=&-39481565540352000 R_{a}{}^{b} R_{b}{}^{a} R_{c}{}^{e} R_{d}{}^{c} \
R_{e}{}^{d} + 496958473622415360 R_{a}{}^{d} R_{b}{}^{a} R_{c}{}^{b} \
R_{d}{}^{c} R\nonumber\\
&- 366085018636185600 R_{a}{}^{c} R_{b}{}^{a} \
R_{c}{}^{b} R^2 + 56771103624384000 R_{a}{}^{b} R_{b}{}^{a} R^3\nonumber\\
&-4236457006581120 R^5 + 605739537316331520 R_{b}{}^{a} R_{d}{}^{b} \
R_{e}{}^{c} R R_{ac}{}^{de}\nonumber\\
&+ 25066678861324800 R^3 R_{ab}{}^{cd} \
R_{cd}{}^{ab} + 2911274422692480 R \left(R_{ab}{}^{cd}{}
R_{cd}{}^{ab}{}\right)^2 \nonumber\\
&- 9235519903334400 R_{a}{}^{b} R_{b}{}^{a} \
R_{e}{}^{c} R_{f}{}^{d} R_{cd}{}^{ef}\nonumber\\
&-654135376602562560 \
R_{c}{}^{a} R_{d}{}^{b} R R_{ab}{}^{ef} R_{ef}{}^{cd}\nonumber\\
&+ \
384078592166215680 R_{b}{}^{a} R_{c}{}^{b} R R_{ad}{}^{ef} \
R_{ef}{}^{cd}\nonumber\\
&- 128301089938030080 R_{a}{}^{b} R_{b}{}^{a} R \
R_{cd}{}^{ef} R_{ef}{}^{cd}\nonumber\\
&+ 247957574993141760 R_{b}{}^{a} R \
R_{ad}{}^{bc} R_{ce}{}^{fg} R_{fg}{}^{de}\nonumber\\
&- 54410152259543040 \
R_{b}{}^{a} R R_{ac}{}^{fg} R_{de}{}^{bc} R_{fg}{}^{de}\nonumber\\
&- \
915942099386695680 R_{ag}{}^{cd} R_{bi}{}^{ef} R_{ce}{}^{ab} \
R_{dj}{}^{gh} R_{fh}{}^{ij} \nonumber\\
&+ 1855713735622656000 R_{ag}{}^{cd} \
R_{bi}{}^{ef} R_{ce}{}^{ab} R_{dh}{}^{ij} R_{fj}{}^{gh}\nonumber\\
&- \
2983978700100403200 R_{ag}{}^{cd} R_{bh}{}^{ij} R_{ce}{}^{ab} \
R_{di}{}^{ef} R_{fj}{}^{gh}\nonumber\\
&- 3268733794665431040 R_{aj}{}^{gh} \
R_{bd}{}^{ij} R_{ce}{}^{ab} R_{fg}{}^{cd} R_{hi}{}^{ef}\, ,\\
\mathcal{S}_{[D=6,n=5]}^{\rm  III}=&-113245541360640 R_{a}{}^{b} R_{b}{}^{a} R_{c}{}^{e} R_{d}{}^{c} \
R_{e}{}^{d} + 1060631652273264 R_{a}{}^{d} R_{b}{}^{a} R_{c}{}^{b} \
R_{d}{}^{c} R\nonumber\\
&- 903602985933600 R_{a}{}^{c} R_{b}{}^{a} R_{c}{}^{b} \
R^2 + 127080097757820 R_{a}{}^{b} R_{b}{}^{a} R^3 - 8955723921633 R^5\nonumber\\
&+ 1791407446201728 R_{b}{}^{a} R_{d}{}^{b} R_{e}{}^{c} R \
R_{ac}{}^{de} + 65583784852200 R^3 R_{ab}{}^{cd} R_{cd}{}^{ab} \nonumber\\
&+17709732531387 R \left(R_{ab}{}^{cd}{} R_{cd}{}^{ab}{}\right)^2 + 3780034053120 \
R_{a}{}^{b} R_{b}{}^{a} R_{e}{}^{c} R_{f}{}^{d} R_{cd}{}^{ef}\nonumber\\
&-\
2136457519124544 R_{c}{}^{a} R_{d}{}^{b} R R_{ab}{}^{ef} \
R_{ef}{}^{cd} + 1548204355449792 R_{b}{}^{a} R_{c}{}^{b} R
R_{ad}{}^{ef} R_{ef}{}^{cd}\nonumber\\
&- 341027462136492 R_{a}{}^{b} R_{b}{}^{a} \
R R_{cd}{}^{ef} R_{ef}{}^{cd} + 601767492758784 R_{b}{}^{a} R \
R_{ad}{}^{bc} R_{ce}{}^{fg} R_{fg}{}^{de}\nonumber\\
&- 195741719323776 
R_{b}{}^{a} R R_{ac}{}^{fg} R_{de}{}^{bc} R_{fg}{}^{de}\nonumber\\
&-
686045879580672 R_{ag}{}^{cd} R_{bi}{}^{ef} R_{ce}{}^{ab} \
R_{dj}{}^{gh} R_{fh}{}^{ij}\nonumber\\
&- 409211547264000 R_{ag}{}^{cd} \
R_{bi}{}^{ef} R_{ce}{}^{ab} R_{dh}{}^{ij} R_{fj}{}^{gh}\nonumber\\
&-4137732154183680 R_{ag}{}^{cd} R_{bh}{}^{ij} R_{ce}{}^{ab} \
R_{di}{}^{ef} R_{fj}{}^{gh}\nonumber\\
&- 8161945395342336 R_{aj}{}^{gh} \
R_{bd}{}^{ij} R_{ce}{}^{ab} R_{fg}{}^{cd} R_{hi}{}^{ef}\, ,
\end{align}
and for them
\begin{align}
\tau_{[D=6,n=5]}^{\rm  I}&=\tau_{(5,5)}-10\tau_{(5,2)}\, ,\\
\tau_{[D=6,n=5]}^{\rm  II}&=\tau_{(5,3)} - 3 \tau_{(5,2)} + \tau_{(5,1)} \, ,\\
\tau_{[D=6,n=5]}^{\rm  III}&=\tau_{(5, 4)} - \tau_{(5, 3)} - 3 \tau_{(5, 2)}\, .
\end{align}

At order six we have four inequivalent GQT classes. Representatives in $D=5$ are given by
\begin{align}
\mathcal{S}_{[D=5,n=6]}^{\rm  I}=&
-73164000 \left(R_{ab}{} R^{ab}{}\right)^3 - 1714893120 R_{ab} R^{ab} 
R_{c}{}^{e} R_{d}{}^{c} R_{e}{}^{d} R \nonumber\\ &+ 1318812172 R_{ab} R^{ab} 
R_{c}{}^{d} R_{d}{}^{c} R^2+ 271196208 R_{a}{}^{c} R_{b}{}^{a} 
R_{c}{}^{b} R^3 \nonumber\\ &- 317404865 R_{ab} R^{ab} R^4 + 18018062 R^6+ 
300979224 R_{c}{}^{a} R_{d}{}^{b} R^3 R_{ab}{}^{cd} \nonumber\\ &+ 248125440 
R_{b}{}^{a} R_{d}{}^{b} R_{e}{}^{c} R^2 R_{ac}{}^{de} + 170805000 
\left(R_{ef}{} R^{ef}{}\right)^2 R_{abcd} R^{abcd} \nonumber\\ & - 452092811 R_{ef} R^{ef} R^2 
R_{abcd} R^{abcd} + 74766829 R^4 R_{abcd} R^{abcd}  \nonumber\\ & - 139080000 R_{ef} 
R^{ef} \left(R_{abcd}{} R^{abcd}{}\right)^2  + 38179125 R^2 \left(R_{abcd}{} 
R^{abcd}{}\right)^2  \nonumber\\ &+ 35080000 \left(R_{abcd}{} R^{abcd}{}\right)^3  - 2244499440 R_{ab} 
R^{ab} R_{e}{}^{c} R_{f}{}^{d} R R_{cd}{}^{ef} \nonumber\\ & - 445474968 
R_{b}{}^{a} R^3 R_{ac}{}^{de} R_{de}{}^{bc}  - 87720000 
\left(R_{a}{}^{c}{}_{b}{}^{d}{} R_{c}{}^{e}{}_{d}{}^{f}{}
R_{e}{}^{a}{}_{f}{}^{b}{}\right)^2  \nonumber\\ & + 84746910 R^3 R_{ab}{}^{ef} 
R_{cd}{}^{ab} R_{ef}{}^{cd} + 2407239480 R_{ab} R^{ab} R_{d}{}^{c} R 
R_{ce}{}^{fg} R_{fg}{}^{de} \nonumber\\ & - 88583040 R_{b}{}^{a} R^2 R_{ad}{}^{bc} 
R_{ce}{}^{fg} R_{fg}{}^{de} - 410141550 R_{ab} R^{ab} R R_{cd}{}^{gh} 
R_{ef}{}^{cd} R_{gh}{}^{ef}  \nonumber\\ & + 564422400 R_{b}{}^{a} R R_{ad}{}^{bc} 
R_{cf}{}^{de} R_{eg}{}^{hi} R_{hi}{}^{fg}\nonumber\\ & - 61305000 R R_{abcd} 
R^{abcd} R_{ef}{}^{ij} R_{gh}{}^{ef} R_{ij}{}^{gh} \nonumber\\ & + 727920000 R^{de} 
R_{abcd} R^{abc}{}_{e} R_{g}{}^{i}{}_{h}{}^{j} 
R_{i}{}^{k}{}_{j}{}^{l} R_{k}{}^{g}{}_{l}{}^{h}\nonumber\\ & - 578160000 
R_{ab}{}^{cd} R_{cd}{}^{ef} R_{ef}{}^{ab} R_{g}{}^{i}{}_{h}{}^{j} 
R_{i}{}^{k}{}_{j}{}^{l} R_{k}{}^{g}{}_{l}{}^{h}\, , \\
\mathcal{S}_{[D=5,n=6]}^{\rm  II}=&-
137140000 \left(R_{ab}{} R^{ab}{}\right)^3 - 1947491520 R_{ab}
R^{ab} R_{c}{}^{e} R_{d}{}^{c} R_{e}{}^{d} R  \nonumber\\& + 1751816692 R_{ab}
R^{ab} R_{c}{}^{d} R_{d}{}^{c} R^2  + 329051088 R_{a}{}^{c}
R_{b}{}^{a} R_{c}{}^{b} R^3 \nonumber\\& - 432438015 R_{ab} R^{ab} R^4 + 25289682
R^6 + 400229864 R_{c}{}^{a} R_{d}{}^{b} R^3 R_{ab}{}^{cd}\nonumber\\&  + 165181440
R_{b}{}^{a} R_{d}{}^{b} R_{e}{}^{c} R^2 R_{ac}{}^{de} + 272619000
\left(R_{ef}{} R^{ef}{}\right)^2 R_{abcd} R^{abcd} \nonumber\\& - 609591221 
R_{ef} R^{ef} R^2 R_{abcd} R^{abcd}  + 100315219 R^4 R_{abcd} R^{abcd} 
\nonumber\\&- 173400000 R_{ef} R^{ef} \left(R_{abcd}{} R^{abcd}{}\right)^2 + 
46512875 R^2 \left(R_{abcd}{} R^{abcd}{}\right)^2\nonumber\\&  + 35600000 
\left(R_{abcd}{} R^{abcd}{}\right)^3 - 2869300240 R_{ab} R^{ab} 
R_{e}{}^{c} R_{f}{}^{d} R R_{cd}{}^{ef} \nonumber\\& - 484473448 R_{b}{}^{a} R^3 
R_{ac}{}^{de} R_{de}{}^{bc}  - 31320000 
\left(R_{a}{}^{c}{}_{b}{}^{d}{} R_{c}{}^{e}{}_{d}{}^{f}{} 
R_{e}{}^{a}{}_{f}{}^{b}{}\right)^2 \nonumber\\& + 55581410 R^3 R_{ab}{}^{ef} 
R_{cd}{}^{ab} R_{ef}{}^{cd} + 2452251080 R_{ab} R^{ab} R_{d}{}^{c} R 
R_{ce}{}^{fg} R_{fg}{}^{de} \nonumber\\& + 118104960 R_{b}{}^{a} R^2 R_{ad}{}^{bc} 
R_{ce}{}^{fg} R_{fg}{}^{de} - 242892050 R_{ab} R^{ab} R R_{cd}{}^{gh} 
R_{ef}{}^{cd} R_{gh}{}^{ef}\nonumber\\&  + 425990400 R_{b}{}^{a} R R_{ad}{}^{bc} 
R_{cf}{}^{de} R_{eg}{}^{hi} R_{hi}{}^{fg} \nonumber\\&- 77575000 R R_{abcd} 
R^{abcd} R_{ef}{}^{ij} R_{gh}{}^{ef} R_{ij}{}^{gh}\nonumber\\&  + 2129040000 
R^{de} R_{abcd} R^{abc}{}_{e} R_{g}{}^{i}{}_{h}{}^{j} 
R_{i}{}^{k}{}_{j}{}^{l} R_{k}{}^{g}{}_{l}{}^{h} \nonumber\\&- 909840000 
R_{ab}{}^{cd} R_{cd}{}^{ef} R_{ef}{}^{ab} R_{g}{}^{i}{}_{h}{}^{j} 
R_{i}{}^{k}{}_{j}{}^{l} R_{k}{}^{g}{}_{l}{}^{h}\, , \\
\mathcal{S}_{[D=5,n=6]}^{\rm  III}=&-859300000 \left(R_{ab}{} R^{ab}{}\right)^3 - 25179802560 R_{ab} R^{ab} 
R_{c}{}^{e} R_{d}{}^{c} R_{e}{}^{d} R \nonumber\\& + 19703296676 R_{ab} R^{ab} 
R_{c}{}^{d} R_{d}{}^{c} R^2 + 4227840144 R_{a}{}^{c} R_{b}{}^{a} 
R_{c}{}^{b} R^3  \nonumber\\& - 4975158595 R_{ab} R^{ab} R^4 + 291039066 R^6 + 
5123673672 R_{c}{}^{a} R_{d}{}^{b} R^3 R_{ab}{}^{cd} \nonumber\\&+ 1331589120 
R_{b}{}^{a} R_{d}{}^{b} R_{e}{}^{c} R^2 R_{ac}{}^{de} + 2222415000 
\left(R_{ef}{} R^{ef}{}\right)^2 R_{abcd} R^{abcd}\nonumber\\& - 6346768033 R_{ef} R^{ef} 
R^2 R_{abcd} R^{abcd} + 1019618087 R^4 R_{abcd} R^{abcd}   \nonumber\\& - 1819320000 
R_{ef} R^{ef} \left(R_{abcd}{} R^{abcd}{}\right)^2  + 513207375 R^2 \left(R_{abcd}{} 
R^{abcd}{}\right)^2 \nonumber\\& + 450800000 \left(R_{abcd}{} R^{abcd}{}\right)^3 - 33156269520 
R_{ab} R^{ab} R_{e}{}^{c} R_{f}{}^{d} R R_{cd}{}^{ef}\nonumber\\&  - 6081896904 
R_{b}{}^{a} R^3 R_{ac}{}^{de} R_{de}{}^{bc} + 4158600000 
\left(R_{a}{}^{c}{}_{b}{}^{d}{} R_{c}{}^{e}{}_{d}{}^{f}{}
R_{e}{}^{a}{}_{f}{}^{b}{}\right)^2\nonumber\\& + 1060299930 R^3 R_{ab}{}^{ef} 
R_{cd}{}^{ab} R_{ef}{}^{cd}  + 34472700840 R_{ab} R^{ab} R_{d}{}^{c} R 
R_{ce}{}^{fg} R_{fg}{}^{de}\nonumber\\& - 834145920 R_{b}{}^{a} R^2 R_{ad}{}^{bc} 
R_{ce}{}^{fg} R_{fg}{}^{de} - 5503384650 R_{ab} R^{ab} R 
R_{cd}{}^{gh} R_{ef}{}^{cd} R_{gh}{}^{ef} \nonumber\\& + 6734419200 R_{b}{}^{a} R 
R_{ad}{}^{bc} R_{cf}{}^{de} R_{eg}{}^{hi} R_{hi}{}^{fg}  \nonumber\\&- 809475000 R 
R_{abcd} R^{abcd} R_{ef}{}^{ij} R_{gh}{}^{ef} R_{ij}{}^{gh}\nonumber\\& + 
15109200000 R^{de} R_{abcd} R^{abc}{}_{e} R_{g}{}^{i}{}_{h}{}^{j} 
R_{i}{}^{k}{}_{j}{}^{l} R_{k}{}^{g}{}_{l}{}^{h}\nonumber\\&  - 9162000000 
R_{ab}{}^{cd} R_{cd}{}^{ef} R_{ef}{}^{ab} R_{g}{}^{i}{}_{h}{}^{j} 
R_{i}{}^{k}{}_{j}{}^{l} R_{k}{}^{g}{}_{l}{}^{h}\, , \\
\mathcal{S}_{[D=5,n=6]}^{\rm  IV}=&+ 31500000 \left(R_{ab}{} R^{ab}{}\right)^3 - 4028310720 R_{ab} R^{ab} 
R_{c}{}^{e} R_{d}{}^{c} R_{e}{}^{d} R  \nonumber\\& + 2252042612 R_{ab} R^{ab} 
R_{c}{}^{d} R_{d}{}^{c} R^2  + 683314128 R_{a}{}^{c} R_{b}{}^{a} 
R_{c}{}^{b} R^3\nonumber\\& - 555694015 R_{ab} R^{ab} R^4 + 27464642 R^6 + 
877183464 R_{c}{}^{a} R_{d}{}^{b} R^3 R_{ab}{}^{cd} \nonumber\\& - 96706560 
R_{b}{}^{a} R_{d}{}^{b} R_{e}{}^{c} R^2 R_{ac}{}^{de} + 163995000 
\left(R_{ef}{} R^{ef}{}\right)^2 R_{abcd} R^{abcd}\nonumber\\& - 407173621 R_{ef} R^{ef} R^2 
R_{abcd} R^{abcd}  + 62048819 R^4 R_{abcd} R^{abcd} \nonumber\\& - 292440000 R_{ef} 
R^{ef} \left(R_{abcd}{} R^{abcd}{}\right)^2 + 28222875 R^2 \left(R_{abcd}{}
R^{abcd}{}\right)^2\nonumber\\&  + 97600000 \left(R_{abcd}{} R^{abcd}{}\right)^3 - 4629108240 R_{ab} 
R^{ab} R_{e}{}^{c} R_{f}{}^{d} R R_{cd}{}^{ef} \nonumber\\& - 1045548648 
R_{b}{}^{a} R^3 R_{ac}{}^{de} R_{de}{}^{bc} + 2409000000 \left(
R_{a}{}^{c}{}_{b}{}^{d}{} R_{c}{}^{e}{}_{d}{}^{f}{}
R_{e}{}^{a}{}_{f}{}^{b}{}\right)^2 \nonumber\\&  + 280021410 R^3 R_{ab}{}^{ef} 
R_{cd}{}^{ab} R_{ef}{}^{cd}  + 6317083080 R_{ab} R^{ab} R_{d}{}^{c} R 
R_{ce}{}^{fg} R_{fg}{}^{de} \nonumber\\&  - 655655040 R_{b}{}^{a} R^2 R_{ad}{}^{bc} 
R_{ce}{}^{fg} R_{fg}{}^{de}  - 1401052050 R_{ab} R^{ab} R 
R_{cd}{}^{gh} R_{ef}{}^{cd} R_{gh}{}^{ef}\nonumber\\&  + 664070400 R_{b}{}^{a} R 
R_{ad}{}^{bc} R_{cf}{}^{de} R_{eg}{}^{hi} R_{hi}{}^{fg} - 46575000 R 
R_{abcd} R^{abcd} R_{ef}{}^{ij} R_{gh}{}^{ef} R_{ij}{}^{gh} \nonumber\\& + 
1414800000 R^{de} R_{abcd} R^{abc}{}_{e} R_{g}{}^{i}{}_{h}{}^{j} 
R_{i}{}^{k}{}_{j}{}^{l} R_{k}{}^{g}{}_{l}{}^{h}\nonumber\\&  - 1832400000 
R_{ab}{}^{cd} R_{cd}{}^{ef} R_{ef}{}^{ab} R_{g}{}^{i}{}_{h}{}^{j} 
R_{i}{}^{k}{}_{j}{}^{l} R_{k}{}^{g}{}_{l}{}^{h}\, .
\end{align}
And the corresponding $\tau(r)$ are given by
\begin{align}
\tau_{[D=5,n=6]}^{\rm  I}&=+\tau_{(6,0)}+12 \tau_{(6,5)}-8 \tau_{(6,6)}\, , \\
\tau_{[D=5,n=6]}^{\rm  II}&=-5 \tau_{(6,2)}-16 \tau_{(6,5)}+11 \tau_{(6,6)}\, , \\
\tau_{[D=5,n=6]}^{\rm III}&=-5\tau_{(6,3)}-3 \tau_{(6,5)}+3 \tau_{(6,6)}\, , \\
\tau_{[D=5,n=6]}^{\rm IV}&=+15 \tau_{(6,4)}-2 (6 \tau_{(6,5)}-\tau_{(6,6)}) \, .
\end{align}


For $D=6$ we find
\begin{align}
\mathcal{S}^{ \rm I}_{[D=6,n=6]}=&-14096679060821760 R_{a}{}^{b} R_{b}{}^{c} R_{c}{}^{d} R_{d}{}^{e} \
R_{e}{}^{f} R_{f}{}^{a}\nonumber\\
&+ 14852647970900544 R_{c}{}^{e} R_{d}{}^{c} \
R_{e}{}^{d} R_{i}{}^{j} R_{j}{}^{i} R\nonumber\\
&- 5617985150718012 \
\left(R_{i}{}^{j}{} R_{j}{}^{i}{}\right)^2 R^2 - 1124843605416416 R_{a}{}^{c} \
R_{b}{}^{a} R_{c}{}^{b} R^3\nonumber\\
&+ 1005726172300248 R_{ab} R^{ab} R^4 - \
29156254184830 R^6 \nonumber\\
&- 1438756007591232 R_{c}{}^{a} R_{d}{}^{b} R^3 \
R_{ab}{}^{cd} + 2380028275859520 R^2 R_{b}{}^{a} R_{d}{}^{b} \
R_{e}{}^{c} R_{ac}{}^{de}\nonumber\\
&+ 1254308457170736 R_{ef} R^{ef} R^2 \
R_{abcd} R^{abcd} - 168004022190642 R^4 R_{ab}{}^{cd} R_{cd}{}^{ab}\nonumber\\
&+ \
3230088574927500 R_{i}{}^{j} R_{j}{}^{i} \left(R_{ab}{}^{cd}{} \
R_{cd}{}^{ab}{}\right)^2 - 607399901908371 R^2 \left(R_{ab}{}^{cd}{}
R_{cd}{}^{ab}{}\right)^2\nonumber\\
&+ 721416483693312 R_{e}{}^{c} R_{f}{}^{d} \
R_{i}{}^{j} R_{j}{}^{i} R R_{cd}{}^{ef}+ 1133891404354368 \
R_{b}{}^{a} R^3 R_{ac}{}^{de} R_{de}{}^{bc} \nonumber\\
&- 682346981951712 R^3 \
R_{ab}{}^{ef} R_{cd}{}^{ab} R_{ef}{}^{cd} + 9376966635379200 R^{ab} \
R^{cd} R_{i}{}^{j} R_{j}{}^{i} R_{ecfd} R^{e}{}_{a}{}^{f}{}_{b} \nonumber\\
&- \
8990642116684800 R^{ab} R_{i}{}^{j} R_{j}{}^{i} \
R_{a}{}^{c}{}_{b}{}^{d} R_{efgc} R^{efg}{}_{d} \nonumber\\
&- 6299359808303232 \
R_{d}{}^{c} R_{i}{}^{j} R_{j}{}^{i} R R_{ce}{}^{fg} R_{fg}{}^{de}\nonumber\\
& + \
1901604108792960 R_{b}{}^{a} R^2 R_{ad}{}^{bc} R_{ce}{}^{fg} \
R_{fg}{}^{de} \nonumber\\
&+ 3847116811602240 R_{i}{}^{j} R_{j}{}^{i} R \
R_{cd}{}^{gh} R_{ef}{}^{cd} R_{gh}{}^{ef} \nonumber\\
&- 10134930764312640 \
R_{a}{}^{b} R_{b}{}^{c} R_{c}{}^{d} R_{d}{}^{a} R_{ef}{}^{hi} \
R_{hi}{}^{eg} \nonumber\\
&- 2178824133657600 R_{b}{}^{a} R R_{ad}{}^{bc} \
R_{cf}{}^{de} R_{eg}{}^{hi} R_{hi}{}^{fg} \nonumber\\
&+ 304956123151680 R \
R_{ab}{}^{cd} R_{cd}{}^{ab} R_{ef}{}^{ij} R_{gh}{}^{ef} R_{ij}{}^{gh} \nonumber\\
&
- 1895257162656000 R_{ab}{}^{cd} R_{cd}{}^{ef} R_{ef}{}^{gh} \
R_{gh}{}^{ij} R_{ij}{}^{kl} R_{kl}{}^{ab} \nonumber\\
&+ 1259726446836000 \
R_{ab}{}^{cd} R_{cd}{}^{ef} R_{ef}{}^{gh} R_{gh}{}^{ab} R_{ij}{}^{kl} \
R_{kl}{}^{ij}\, ,\\
\mathcal{S}_{[D=6,n=6]}^{\rm II}=&-31836692340236160 R_{a}{}^{b} R_{b}{}^{c} R_{c}{}^{d} R_{d}{}^{e} \
R_{e}{}^{f} R_{f}{}^{a}\nonumber\\
&+ 34439506371202464 R_{c}{}^{e} R_{d}{}^{c} \
R_{e}{}^{d} R_{i}{}^{j} R_{j}{}^{i} R - 13557698416858564 \
\left(R_{i}{}^{j}{} R_{j}{}^{i}{}\right)^2 R^2 \nonumber\\
&- 2849781769779440 R_{a}{}^{c} \
R_{b}{}^{a} R_{c}{}^{b} R^3 + 2611991351109630 R_{ab} R^{ab} R^4\nonumber\\
& - \
88399029128845 R^6 - 3677104626840832 R_{c}{}^{a} R_{d}{}^{b} R^3 \
R_{ab}{}^{cd} \nonumber\\
&+ 6132894365769600 R^2 R_{b}{}^{a} R_{d}{}^{b} \
R_{e}{}^{c} R_{ac}{}^{de} + 3157451844617752 R_{ef} R^{ef} R^2 \
R_{abcd} R^{abcd} \nonumber\\
&- 441312471667562 R^4 R_{ab}{}^{cd} R_{cd}{}^{ab} + \
7146998363226150 R_{i}{}^{j} R_{j}{}^{i} \left(R_{ab}{}^{cd}{} \
R_{cd}{}^{ab}{}\right)^2 \nonumber\\
&- 1314167139538110 R^2 \left(R_{ab}{}^{cd}{} \
R_{cd}{}^{ab}{}\right)^2 + 2087392939560192 R_{e}{}^{c} R_{f}{}^{d} \
R_{i}{}^{j} R_{j}{}^{i} R R_{cd}{}^{ef} \nonumber\\
&+ 2787439490093632 \
R_{b}{}^{a} R^3 R_{ac}{}^{de} R_{de}{}^{bc} - 1533575360560320 R^3 \
R_{ab}{}^{ef} R_{cd}{}^{ab} R_{ef}{}^{cd} \nonumber\\
&+ 22373284326307200 R^{ab} \
R^{cd} R_{i}{}^{j} R_{j}{}^{i} R_{ecfd} R^{e}{}_{a}{}^{f}{}_{b} \nonumber\\
&- \
20005506406828800 R^{ab} R_{i}{}^{j} R_{j}{}^{i} \
R_{a}{}^{c}{}_{b}{}^{d} R_{efgc} R^{efg}{}_{d} \nonumber\\
&- 15079987603900032 \
R_{d}{}^{c} R_{i}{}^{j} R_{j}{}^{i} R R_{ce}{}^{fg} R_{fg}{}^{de} \nonumber\\
&+ \
3611210786726400 R_{b}{}^{a} R^2 R_{ad}{}^{bc} R_{ce}{}^{fg} \
R_{fg}{}^{de}\nonumber\\
&+ 8724416327193600 R_{i}{}^{j} R_{j}{}^{i} R \
R_{cd}{}^{gh} R_{ef}{}^{cd} R_{gh}{}^{ef}\nonumber\\
& - 23212861724463840 \
R_{a}{}^{b} R_{b}{}^{c} R_{c}{}^{d} R_{d}{}^{a} R_{ef}{}^{hi} \
R_{hi}{}^{eg}\nonumber\\
&- 3819771274176000 R_{b}{}^{a} R R_{ad}{}^{bc} \
R_{cf}{}^{de} R_{eg}{}^{hi} R_{hi}{}^{fg}\nonumber\\
&+ 616693394116800 R \
R_{ab}{}^{cd} R_{cd}{}^{ab} R_{ef}{}^{ij} R_{gh}{}^{ef} R_{ij}{}^{gh} \
\nonumber\\
&- 4293178347744000 R_{ab}{}^{cd} R_{cd}{}^{ef} R_{ef}{}^{gh} \
R_{gh}{}^{ij} R_{ij}{}^{kl} R_{kl}{}^{ab} \nonumber\\
&+ 2871990255420000 \
R_{ab}{}^{cd} R_{cd}{}^{ef} R_{ef}{}^{gh} R_{gh}{}^{ab} R_{ij}{}^{kl} \
R_{kl}{}^{ij}\, ,\\
\mathcal{S}^{\rm III}_{[D=6,n=6]}=&-6943970290757760 R_{a}{}^{b} R_{b}{}^{c} R_{c}{}^{d} R_{d}{}^{e} \
R_{e}{}^{f} R_{f}{}^{a}\nonumber\\
&+ 5474732611842144 R_{c}{}^{e} R_{d}{}^{c} \
R_{e}{}^{d} R_{i}{}^{j} R_{j}{}^{i} R - 2281174181020312 \
\left(R_{i}{}^{j}{} R_{j}{}^{i}{}\right)^2 R^2\nonumber\\
&- 355850581360016 R_{a}{}^{c} \
R_{b}{}^{a} R_{c}{}^{b} R^3 + 372382664514798 R_{ab} R^{ab} R^4 - \
8079133402255 R^6 \nonumber\\
&- 724084260087232 R_{c}{}^{a} R_{d}{}^{b} R^3 \
R_{ab}{}^{cd} + 680829460259520 R^2 R_{b}{}^{a} R_{d}{}^{b} \
R_{e}{}^{c} R_{ac}{}^{de} \nonumber\\
&+ 397223834424736 R_{ef} R^{ef} R^2 \
R_{abcd} R^{abcd} - 65621870854892 R^4 R_{ab}{}^{cd} R_{cd}{}^{ab} \nonumber\\
&
+656091001244250 R_{i}{}^{j} R_{j}{}^{i} \left(R_{ab}{}^{cd}{} \
R_{cd}{}^{ab}{}\right)^2 - 121001538886371 R^2 \left(R_{ab}{}^{cd}{} \
R_{cd}{}^{ab}{}\right)^2 \nonumber\\
&+ 2073302769914112 R_{e}{}^{c} R_{f}{}^{d} \
R_{i}{}^{j} R_{j}{}^{i} R R_{cd}{}^{ef} + 671781101071168 R_{b}{}^{a} \
R^3 R_{ac}{}^{de} R_{de}{}^{bc}\nonumber\\
&- 224552737043712 R^3 R_{ab}{}^{ef} \
R_{cd}{}^{ab} R_{ef}{}^{cd}\nonumber\\
&+ 2768431158979200 R^{ab} R^{cd} \
R_{i}{}^{j} R_{j}{}^{i} R_{ecfd} R^{e}{}_{a}{}^{f}{}_{b}\nonumber\\
&- \
2668237319404800 R^{ab} R_{i}{}^{j} R_{j}{}^{i} \
R_{a}{}^{c}{}_{b}{}^{d} R_{efgc} R^{efg}{}_{d}\nonumber\\
&- 3774533321404032 \
R_{d}{}^{c} R_{i}{}^{j} R_{j}{}^{i} R R_{ce}{}^{fg} R_{fg}{}^{de}\nonumber\\
&+ \
523902114552960 R_{b}{}^{a} R^2 R_{ad}{}^{bc} R_{ce}{}^{fg} \
R_{fg}{}^{de}\nonumber\\
&+ 1290427980114240 R_{i}{}^{j} R_{j}{}^{i} R \
R_{cd}{}^{gh} R_{ef}{}^{cd} R_{gh}{}^{ef}\nonumber\\
&- 849014886788640 \
R_{a}{}^{b} R_{b}{}^{c} R_{c}{}^{d} R_{d}{}^{a} R_{ef}{}^{hi} \
R_{hi}{}^{eg}\nonumber\\
&- 965396466777600 R_{b}{}^{a} R R_{ad}{}^{bc} \
R_{cf}{}^{de} R_{eg}{}^{hi} R_{hi}{}^{fg} \nonumber\\
&+ 70932032851680 R \
R_{ab}{}^{cd} R_{cd}{}^{ab} R_{ef}{}^{ij} R_{gh}{}^{ef} R_{ij}{}^{gh} \nonumber\\
&
- 336455941536000 R_{ab}{}^{cd} R_{cd}{}^{ef} R_{ef}{}^{gh} \
R_{gh}{}^{ij} R_{ij}{}^{kl} R_{kl}{}^{ab}\nonumber\\
& + 219456058536000 \
R_{ab}{}^{cd} R_{cd}{}^{ef} R_{ef}{}^{gh} R_{gh}{}^{ab} R_{ij}{}^{kl} \
R_{kl}{}^{ij} \, ,\\
\mathcal{S}^{\rm IV}_{[D=6,n=6]}=&-14096679060821760 R_{a}{}^{b} R_{b}{}^{c} R_{c}{}^{d} R_{d}{}^{e} \
R_{e}{}^{f} R_{f}{}^{a}\nonumber\\
&+ 14852647970900544 R_{c}{}^{e} R_{d}{}^{c} \
R_{e}{}^{d} R_{i}{}^{j} R_{j}{}^{i} R - 5617985150718012 \
\left(R_{i}{}^{j}{} R_{j}{}^{i}{}\right)^2 R^2\nonumber\\
&- 1124843605416416 R_{a}{}^{c} \
R_{b}{}^{a} R_{c}{}^{b} R^3 + 1005726172300248 R_{ab} R^{ab} R^4 \nonumber\\
&- \
29156254184830 R^6 - 1438756007591232 R_{c}{}^{a} R_{d}{}^{b} R^3 \
R_{ab}{}^{cd}\nonumber\\
&+ 2380028275859520 R^2 R_{b}{}^{a} R_{d}{}^{b} \
R_{e}{}^{c} R_{ac}{}^{de} + 1254308457170736 R_{ef} R^{ef} R^2 \
R_{abcd} R^{abcd}\nonumber\\
&- 168004022190642 R^4 R_{ab}{}^{cd} R_{cd}{}^{ab} + \
3230088574927500 R_{i}{}^{j} R_{j}{}^{i} \left(R_{ab}{}^{cd}{}
R_{cd}{}^{ab}{}\right)^2 \nonumber\\
&- 607399901908371 R^2 \left(R_{ab}{}^{cd}{} \
R_{cd}{}^{ab}{}\right)^2 + 721416483693312 R_{e}{}^{c} R_{f}{}^{d} \
R_{i}{}^{j} R_{j}{}^{i} R R_{cd}{}^{ef}\nonumber\\
&+ 1133891404354368 \
R_{b}{}^{a} R^3 R_{ac}{}^{de} R_{de}{}^{bc}- 682346981951712 R^3 \
R_{ab}{}^{ef} R_{cd}{}^{ab} R_{ef}{}^{cd} \nonumber\\
&+ 9376966635379200 R^{ab} \
R^{cd} R_{i}{}^{j} R_{j}{}^{i} R_{ecfd} R^{e}{}_{a}{}^{f}{}_{b}- \nonumber\\
&\
8990642116684800 R^{ab} R_{i}{}^{j} R_{j}{}^{i} \
R_{a}{}^{c}{}_{b}{}^{d} R_{efgc} R^{efg}{}_{d} \nonumber\\
&- 6299359808303232 \
R_{d}{}^{c} R_{i}{}^{j} R_{j}{}^{i} R R_{ce}{}^{fg} R_{fg}{}^{de} \nonumber\\
&+ \
1901604108792960 R_{b}{}^{a} R^2 R_{ad}{}^{bc} R_{ce}{}^{fg} \
R_{fg}{}^{de}\nonumber\\
& + 3847116811602240 R_{i}{}^{j} R_{j}{}^{i} R \
R_{cd}{}^{gh} R_{ef}{}^{cd} R_{gh}{}^{ef} \nonumber\\
&- 10134930764312640 \
R_{a}{}^{b} R_{b}{}^{c} R_{c}{}^{d} R_{d}{}^{a} R_{ef}{}^{hi} \
R_{hi}{}^{eg} \nonumber\\
&- 2178824133657600 R_{b}{}^{a} R R_{ad}{}^{bc} \
R_{cf}{}^{de} R_{eg}{}^{hi} R_{hi}{}^{fg} \nonumber\\
&+ 304956123151680 R \
R_{ab}{}^{cd} R_{cd}{}^{ab} R_{ef}{}^{ij} R_{gh}{}^{ef} R_{ij}{}^{gh} \
\nonumber\\
&- 1895257162656000 R_{ab}{}^{cd} R_{cd}{}^{ef} R_{ef}{}^{gh} \
R_{gh}{}^{ij} R_{ij}{}^{kl} R_{kl}{}^{ab}\nonumber\\
& + 1259726446836000 \
R_{ab}{}^{cd} R_{cd}{}^{ef} R_{ef}{}^{gh} R_{gh}{}^{ab} R_{ij}{}^{kl} \
R_{kl}{}^{ij}\, ,
\end{align}
and for them
\begin{align}
\tau_{[D=6,n=6]}^{\rm  I}&=\tau_{(6, 6)} - 15 \tau_{(6, 2)} + 4 \tau_{(6, 1)}\, ,\\
\tau_{[D=6,n=6]}^{\rm  II}&=\tau_{(6, 5)} - 10 \tau_{(6, 2)} + 3 \tau_{(6, 1)}\, ,\\
\tau_{[D=6,n=6]}^{\rm  III}&=\tau_{(6, 4)} - 6 \tau_{(6, 2)} + 2 \tau_{(6, 1)}\, ,\\
\tau_{[D=6,n=6]}^{\rm  IV}&=\tau_{(6, 3)} - 3 \tau_{(6, 2)} + \tau_{(6, 1)}\, .
\end{align}

\bibliographystyle{JHEP}
\bibliography{Gravities}

\end{document}